\documentclass[letterpaper,hyper]{JHEP3}
\usepackage{graphicx}
\usepackage{epsfig,verbatim}
\usepackage{slashed}
\usepackage{amsmath}
\usepackage{amsfonts}
\usepackage{amssymb}
\usepackage{verbatim} 

\newcommand{\beq}{\begin{equation}}
\newcommand{\eeq}{\end{equation}}
\newcommand{\bea}{\begin{eqnarray}}
\newcommand{\eea}{\end{eqnarray}}

\newcommand{\del}{\partial}

\newcommand{\lsim}{\mbox{\raisebox{-.6ex}{~$\stackrel{<}{\sim}$~}}}

\def\sss{{\scriptscriptstyle}}

\title{Warped Radion Inflation}
\author{Joel Trudeau, James M. Cline \\
Department of Physics, McGill University,
 Montr\'{e}al, QC, H3A 2T8, Canada
\email{trudeau@physics.mcgill.ca, jcline@physics.mcgill.ca}}
\preprint{}

\abstract{We show that the radion in a warped geometry 
bounded by two branes can have a potential suitable for inflation.  Our
construction is based upon a solution known in string theory
as the linear dilaton, in which the back-reaction from a bulk scalar $\Phi$ is
exactly accounted for. The radion, stabilized by $\Phi$, is much heavier than
the TeV scale and its couplings to the standard model are much more suppressed
than in the usual Randall-Sundrum solution.  We present a new formalism for obtaining
approximate time-dependent  solutions, based on perturbing the exact solution
to the coupled Einstein and scalar field equations in the bulk.  It allows the
radion potential to be computed directly in terms of the brane potentials for
$\Phi$.   We show that simple exponential potentials on the branes can lead to
a 4D radion potential with a flattened hilltop form, yielding inflation with a
spectral index of typically $n_s=0.96$ and no higher than $0.99$.
With more complicated brane potentials,
the descent from the hilltop can be a linear potential, giving a
tensor-to-scalar ratio as large as $r=0.07$ with $n_s=0.974$.  The couplings of
the radion to the standard model particles are dictated by general covariance,
so the details of reheating are explicitly calculable, leading to a
reheat temperature of at least $10^7$ GeV. The quantum
corrections to the inflaton potential from its couplings to matter
are also calculable and are shown to be
small, so that the prediction for the shape of the potential is under
theoretical control, even with superPlanckian field excursions.}

\begin{document}

\section{Introduction}

The possibility that we live on a brane embedded in a
higher-dimensional space has been widely explored since one of 
its earliest string theory manifestations, the Horava-Witten
model \cite{HW}.  
Brane-worlds afford intriguing opportunities to go beyond the standard
model of particle physics and cosmology, but they also present
challenges for reproducing known physics. The existence of moduli in
these models is a generic feature with potentially important
cosmological consequences.  If unstabilized, they  can conflict with
big bang nucleosynthesis or  fifth force constraints. This was the
case in the original proposal of Randall and Sundrum (RS1) \cite{RS1},
which elegantly addressed the weak-scale hierarchy problem. These issues were resolved by introducing a bulk
scalar field to stabilize the radion, the modulus determining the size of the
extra dimension \cite{GW2}.

Moduli are natural candidates to serve as the inflaton, and the radion
is foremost among them.  However it has been difficult to make radion
inflation work in practice.  Ref.\ \cite{rapid} proposed the modulus
of large extra dimensions \cite{ADD} as the inflaton, but this
required an inflaton potential with a rather peculiar shape to have
inflation at the TeV scale \cite{jcextra}.  In the framework of warped
extra dimensions, most of the effort has been either on inflation
driven by motion of branes within the bulk, or else conventional
inflation on one of the branes modified by an unconventional Friedmann
equation due to the extra dimensions.  We are not aware of previous
proposals for getting inflation from the radion in the RS-I (orbifold
bounded by two branes) scenario.\footnote{a recent reference using
different moduli fields is \cite{sundrum}.}  It would be difficult to
make this work, first because the Goldberger-Wise mechanism for
stabilizing the radion does not give it a flat potential, and secondly
because at Planck-scale values of the canonically normalized radion
field, which could in principle have given chaotic inflation, the
radion has already passed the barrier in the potential separating the
compactified and decompactified regimes of the theory.  In fact the
light KK modes of the extra dimension become excited, invalidating the
4D effective description, long before the radion can reach the Planck
scale.

In this work we present a model of a warped 5D radion as inflaton that
overcomes these difficulties.  We incorporate a warped background
solution that has negative curvature but is not AdS; instead it is
conformally flat, and relies on a linearly varying bulk scalar field  
for the source of its negative bulk energy density.  The weak scale
hierarchy problem is naturally solved, and inflation occurs at a
high scale, with weak couplings of the inflaton to matter on the 
standard model brane.  We are able to relate the potential of the
radion to the bulk scalar field potentials on the branes in a rather
simple way, and to find radion potentials (starting from brane
potentials of exponential form) that are flat enough for inflation
with a certain amount of tuning of parameters of the brane
potentials.  

Generically, one always expects decompactified flat space to be a
solution to the Einstein equations, and so the radion potential
naturally has degenerate minima representing compact and noncompact
fifth dimension, separated by a barrier.  In our model the barrier is
at Planckian field values, and we can find examples where
superPlanckian excursions during inflation lead to an observable
tensor component in the perturbations.   Other examples have a small
tensor component with inflation ending close to the hilltop. Because
the minima are degenerate, we have the situation of topological
inflation \cite{Vilenkin,Linde:1994wt}: the universe divides into 4D and 5D domains separated by 
eternally inflating domain walls where the radion is near the top of
the potential.  This guarantees the existence of regions where
the inflaton is arbitrarily close to the potential maximum (up to the
limitation imposed by quantum fluctuations of the field) so that one
does not have to rely upon fine-tuning of the initial position of the
inflaton to get enough inflation.  

In section \ref{sect2} we introduce the static solution with stabilized
radion and contrast it to the RS solution, especially with regard to
the hierarchy problem.  In section \ref{sect3} we perturb away from the
static solution to find time-dependent ones, in the approximation that
the back-reaction is small.  We derive the effective 4D theory in
section \ref{sect4}, computing the mass of the radion and the spectrum of
Kaluza-Klein excitations for the radion and bulk scalar.  We then
construct some inflationary models using exponential brane potentials
and work out their predictions for the CMB power spectrum, using both
analytic and numerical techniques.  Section \ref{sect5} derives the
couplings of the radion to the standard model, enabling us to compute
the reheat temperature at the end of inflation and to estimate the
radiative corrections to the inflaton potential.  We give conclusions
in section \ref{sect6} and the bulk Einstein and scalar field equations
in the appendix.

\section{Unperturbed static solution}
\label{sect2}

We begin by introducing the framework for our 5D brane-world model.
The action for 5D gravity coupled to the stabilizing
scalar field, $\Phi =\Phi \left( y,t\right) $ with a bulk potential $V\left(
\Phi \right) $ and potentials on the branes $V_{0,1}\left( \Phi \right) $ is 
\bea
  S &= 2 \int_0^{1} d^{\,5}x\, \sqrt{g} \left(- \frac{1}{2\kappa_5^2}R + \frac{1}{2}\del^{\mu}\Phi\del_{\mu}\Phi -V(\Phi) \right)
  -\int d^{\,4}x\, \sqrt{g_4} \left. V_0\right|_{y=0} -\int
d^{\,4}x\,\sqrt{g_4} \left. V_1\right|_{y=1} \nonumber\\
  &\quad -\kappa_5^2\int d^{\,4}x\,\sqrt{g_4}  \left. \left[K\right]\right|_{y=0}
  -\kappa_5^2\int d^{\,4}x\,\sqrt{g_4}  \left. \left[K\right]\right|_{y=1}  \label{action}
\eea
The extra dimension is an $S^{1}/Z_{2}$ orbifold (hence the factor of
$2$) and all functions are
symmetric under $y\rightarrow -y$. Thus we consider the interval $y\in \left[
0,1\right] $ for the compact extra dimension with the Planck $\left(
y_{0}=0\right) $ and TeV $\left( y_{1}=1\right) $ branes located at the
orbifold fixed points. We also include extrinsic curvature terms $[K]$
that are needed for properly defining the variations of the action. 

The most general metric ansatz that respects 3D
homogeneity and isotropy can be written as
\begin{equation}
  ds^{2}=e^{2N\left( t,y\right) }dt^{2}-e^{2A\left( t,y\right) }d\mathbf{x}
  ^{2}-e^{2B\left( t,y\right) }dy^{2}  \label{metric_general}
\end{equation}
The field equations for this ansatz are well known and we
reproduce them in appendix \ref{appA}.

\subsection{Static solutions}

It will be useful for what follows to construct an exact solution
to the coupled bulk scalar field and Einstein equations, which 
is conformally flat:
\beq
  ds^{2} =e^{2n\left( y\right) }\left( dt^{2}-d\mathbf{x}^{2}-b^{2}dy^{2}\right)  \label{metric_stat}
\eeq
Notice that this differs from AdS$_5$ in Randall-Sundrum coordinates
where the $e^{2n\left( y\right) }$ factor does not multiply
$dy^2$.  The negative vacuum energy density provided by the bulk
scalar potential is not constant, so the geometry is not AdS. 
This solution is well-known (see \textit{e.g.,}
\cite{myers}-\cite{polchinski1}) in the context of strings propagating
in a spacetime with subcritical dimensionality, compensated by a
spatially linearly varying dilaton similar to our bulk scalar.  
With this ansatz the Einstein equations in the bulk reduce to 
\begin{align}
  -3\left( n^{\prime \prime }+n^{^{\prime }2}\right) & =\kappa_5^{2}\left( 
  \frac{1}{2}\Phi ^{\prime 2}+b^{2}e^{2n}\,V\left( \Phi \right) \right)
  \label{EE_static1} \\
  -6\left( n^{^{\prime }}\right) ^{2}& =\kappa_5^{2}\left( -\frac{1}{2}\Phi
  ^{\prime 2}+b^{2}e^{2n}\,V\left( \Phi \right) \right)  
  \label{EE_static2}
\end{align}
As usual, the scalar field equation of motion
\beq
  \Phi ^{\prime \prime }+3n^{\prime }\Phi ^{\prime }-b^{2}e^{2n}\frac{dV}{%
  d\Phi }=0  \label{scalar_eq_stat}
\eeq
is not independent of the Einstein equations, but can be obtained
by taking linear combinations of (\ref{EE_static1}, \ref{EE_static2})
and derivatives thereof.  
By choosing the bulk potential to take a special form, $V=
-\bar V \exp(c_{\sss V}\Phi)$, one finds that the solutions for $n$ and $\Phi$ are simple linear
functions of $y$: $n(y)=n_0-kby$, 
$\Phi(y) = \Phi_0 + c_{\sss \Phi}y$.
Here $k$ plays the role of the AdS curvature scale, as in the usual 
RS solution, and by analogy we will so refer to it, even though
the solution is not AdS.  
The integration constants $n_0$ and $\Phi_0$ can be absorbed into
the normalizations of $b$ and $\bar V$ respectively without loss of generality, 
so we set them to zero.  The equations of motion then imply that
$\bar V = -{9 k^2\over 2\kappa_5^2}$, 
$c_{\sss V} = 2{\kappa_5\over\sqrt{3}}$,
$c_{\sss\Phi} = {\sqrt{3}kb\over\kappa_5}$.
With these choices, the bulk scalar potential is given by
\beq
	V(\Phi) =  -{3\over 2} \mu_5^2 k^2
	\exp\left(2\mu_5^{-1}\Phi\right)
\label{bulk_pot_2}
\eeq
where we define
\beq
	\mu_5 = {\sqrt{3}\over \kappa_5}
\eeq
and the solution is
\beq
	n(y) = -kby;\quad \Phi = \mu_5 k b y
\eeq
In contrast to the original RS construction, we do not include a bulk
cosmological constant; the negative bulk potential $V(\Phi)$
is responsible for the curvature of the 5D geometry.  Similar negative
potentials have been studied in the context of Golberger-Wise stabilization
of bulk fields in AdS$_5$ \cite{dewolfe}, where it was observed that the 
unboundedness of the potential does not lead to instabilities in the AdS
background \cite{BF}.  The present case is similar; in the analogous subcritical
string theory situation where the bulk scalar is the dilaton, in the string frame the 
negative coefficient of the potential is interpreted as a negative
bulk cosmological constant, which would give rise to an AdS background if
the dilaton were held fixed.  In the present situation we have in addition
the couplings of the bulk field to the branes, which prevent the field from
running away.

\subsection{Junction conditions}

The Israel junction conditions (see {\it e.g.,} ref.\ \cite{KO}) and the
boundary conditions for the scalar field are given by
\beq
  \left.b^{-1}e^{-n} n^{\prime }\right|_{y_{i}-\epsilon}^{y_{i}+\epsilon}
  =\left.\pm \frac{\kappa_5^{2}V_{i}}{%
  3}\right|_{y_{i}},\quad
  b^{-1}e^{-n}\left. \Phi ^{\prime }\right|_{y_{i}-\epsilon}^{y_{i}+\epsilon} =
  \left. \pm \frac{\partial V_{i}
  }{\partial \Phi }\right|_{y_{i}}  \label{junction2}
\eeq
where the $\pm$ apply respectively at $y_0 = 0$ and $y_1=1$.
Then explicitly,
\begin{align}
  2 \mu_5^2 k & = V_{0}(0),\qquad\quad
	  2 \mu_5^2 k  = -\,e^{n_{1}}\,V_{1}(\Phi_1) 
  \label{junction1_stat} \\
  2 \mu_5 k& = \frac{\partial V_{0}}{\partial
  \Phi }(0),\quad\quad\ 
  2 \mu_5 k = -\,e^{n_{1}}\,\frac{\partial V_{1}}{\partial
  \Phi }(\Phi_1)  \label{junction2_stat}
\end{align}
where $e^{n_{1}}=e^{-kb}$ and $\Phi_1 = \mu_5 kb$
are respectively the warp factor and the scalar field value
on the TeV brane.

Let us recall the physical significance of the junction conditions.
In the pure RS model with no scalar field, the two junction
conditions require fine-tuning of the brane tensions (here
represented by the values of the potentials $V_0$ and $V_1$ on the
branes).  For  generic values of the tensions, one would obtain a
bulk solution that is not static, and which contains a black hole in
the extra dimension \cite{Kaloper:1999sm}.  Thus the two tunings can be interpreted as (1)
the usual setting of the 4D cosmological constant to zero, to obtain
a static solution, and (2) the tuning of the bulk black hole mass to
zero.  The radion $b$ is exactly massless in this solution, and can
thus take any value.

Next consider the addition of the bulk scalar field.  In the usual
implementations of the Goldberger-Wise mechanism, the back-reaction of
the scalar field on the metric is taken to be small, and one
simply solves the bulk scalar equation in the AdS$_5$ background.
This is a second order equation, and the two constants of integration
in the solution are determined by the two additional boundary
conditions involving $dV_i/d\Phi$.  This generically induces a
potential for the radion that is minimized at some value $b_0$.  

In our solution, the situation is somewhat different; we have singled
out a particularly simple form of the scalar field solution, which
is only compatible with certain choices of $dV_i/d\Phi$ at the
boundaries.  It is still true that the radion is stabilized as in 
the generic GW mechanism, but there is an additional tuning of
potential parameters needed to maintain the linear solution for
$\Phi$.  This extra tuning is on the same footing as that of the
bulk black hole mass in the pure RS solution.  We suspect that our
results can be generalized to more complicated solutions in which 
this tuning is relaxed, but for ease of computation, we will adhere
to this special situation.  

The upshot is that the junction conditions amount to three tunings
of brane potential parameters, plus one condition that fixes the
value of the radion $b$, and hence the warp factor on the TeV 
brane.  Let us illustrate with a simple example of brane potentials 
which is similar to that made by GW:
\beq
	V_0 = m^2_0(\Phi + u_0)^2,\quad 
	V_1 = -m^2_1(\Phi+ u_1)^2
\eeq
Solving the junction conditions, we obtain three constraints
that can be regarded as fine-tunings on the parameters 
$m_0$, $u_0$ and $u_1$,
\beq
	m_0^2 = {k\over 2},\quad
	u_0 = 2\mu_5,\quad
	u_1 = {\mu_5}\left(2-kb\right)
\eeq
while $m_1$ can be regarded as adjustable, and its value determines
the warp factor through
\beq
	e^{-kb_0} = {m_0^2\over m_1^2}
\eeq
One notices a disadvantage of our approach relative to the usual one;
we have had to build the hierarchy of scales into our original choice
of brane couplings to get an exponentially small warp factor, as
opposed to having generic Planck-scale values for the brane potential
parameters.  The next example shows that this problem can be ameliorated
if the potentials take an exponential form.

\subsection{Exponential brane potentials}

In our subsequent construction of an inflationary solution, we will
make use of a different choice of brane potentials, which are
exponentials in $\Phi$:
\begin{align}
V_{0}& =\phantom{-}\Lambda _{0}\,e^{\alpha_{0}\Phi/\mu_5 }
-\Delta_{0}\,e^{\beta_{0}\Phi/\mu_5 }  \label{brane_pot1} \\
V_{1}& =-\Lambda _{1}\,e^{\alpha_{1}\Phi/\mu_5 }
+\Delta_{1}\,e^{\beta_{1}\Phi /\mu_5}  \label{brane_pot2}
\end{align}
The junction conditions lead to 
\bea
\Lambda_{0}&=&2\mu_5^2k\left( \frac{1-\beta_{0}}{\alpha
_{0}-\beta _{0}}\right),\quad\qquad\qquad 
\Delta _{0} = 2\mu_5^2k\left( \frac{1-\alpha _{0}}{\alpha
_{0}-\beta _{0}}\right)  \label{0brane_pot_param0}\\
\Lambda_{1}&=&2\mu_5^2k\left( \frac{1-\beta_{1}}{\alpha
_{1}-\beta _{1}}\right)e^{(1-\alpha_1)kb_0},\quad 
\Delta _{1} = 2\mu_5^2k\left( \frac{1-\alpha _{1}}{\alpha
_{1}-\beta _{1}}\right)e^{(1-\beta_1)kb_0}  \label{0brane_pot_param1}
\eea
For a given desired value of $b_0$ determining the hierarchy of scales
between the two branes, there
are thus four free parameters $\alpha_i$, $\beta_i$ that can
eventually be used to tune the potential of the radion to a form
suitable for inflation, while the dimensionful couplings $\Delta_i$
and $\Lambda_i$ are fixed in terms of these.  Notice that if
$\alpha_1$ and $\beta_1$ happen to be moderately close to 1 
({\it e.g.,} $\sim 1.1$), the
explicit hierarchy can be ameliorated between the dimensionful 
parameters on the Planck brane versus those on the TeV, while
yielding a sufficiently large value of $kb_0\sim 37$ to explain the
TeV scale, since then $\Lambda_1/\Lambda_0\sim e^{(1-\alpha_1)kb_0}$
and $\Delta_1/\Delta_0\sim e^{(1-\beta_1)kb_0}$.

\section{Time-dependent solutions}
\label{sect3}

Our goal now is to extend the static solution of the previous section
to dynamical ones in which the radion is displaced from its stable
equilibrium value $b_0$. In refs.\ \cite{KO},\cite{JCHF}-\cite{KT2}, 5D solutions that
also include the
dynamics of a bulk scalar field were found for the case of a single brane.
In these papers the search for solutions was simplified by assuming that
the metric functions and the bulk scalar were separable functions of
$t$ and $y$, which we will justify.
The key to our approach
will be to allow for a large excursion of the radion field 
(which plays the role of the inflaton) in response to linear
perturbations of the metric and bulk scalar.  Although such an
ansatz is not guaranteed to be consistent, we will show that for
suitable choices of the parameters for the exponential 
brane potentials (\ref{brane_pot1},\ref{brane_pot2})
it is justified, and will allow us to find physically
interesting inflationary solutions.

\subsection{Small perturbations with large radion fluctuations}

We make an ansatz for general scalar perturbations similar to that
in ref.\ \cite{CGK}, 
\bea
  ds^{2}&=& e^{2n(y)}\left[e^{2F(x,y)}\eta _{\mu \nu }dx^{\mu}dx^{\nu }
  -b_{0}^{2}\,e^{2\varphi(t) +2G(x,y) }dy^{2}\right]   \label{metricKK_aug}\\
  \Phi(x,y) &=&  -\mu_5\left(n(y) +\varphi(t)\right)+ \delta\Phi(x,y)
  \label{scalarKK_aug}
\eea
It reduces to the static solution of the previous section when
$F = G = \delta\Phi = \varphi(t) = 0$.  Normally then, one would
linearize in all of these quantities.  However we will find a more
general solution for the perturbations by treating only
$F,\,G,\,\delta\Phi$ as being small, while working to all orders in 
$\varphi(t)$ (and $\dot\varphi$).   This corresponds to allowing for large excursions of
the radion during inflation.   We will justify it {\it a postieriori}
by choosing special values of the brane potentials that lead to
a sufficiently flat inflaton potential.

The perturbations $F,\,G,\,\delta\Phi$ are constrained by the 
nondynamical Einstein equations, namely those containing only first
derivatives. The $G_{05}$ equation gives
\beq
  (\dot G  + \mu_5^{-1}\dot{\delta\Phi})n' +  (F'  + \mu_5^{-1}\delta\Phi')\dot\varphi -\dot F' =0  \label{05_derived}
\eeq
Similarly the off-diagonal $G_{ij}$ equation gives
\beq
  2\del_i\del_j F + \del_i\del_j G =0	\label{ij_derived}
\eeq
It can be integrated to find
\beq
  G(x,y) = -2F(x,y) + G_0(t,y) \label{G_pert}
\eeq
where an untenable term linear in $\mathbf{x}$ has been set to zero.
The $G_{0j}$ equation
\beq
  (\del_j F + \mu_5^{-1}\del_j \delta\Phi)\dot\varphi =0	\label{0j_derived}
\eeq
can be solved for the bulk scalar perturbations,
using eq. (\ref{G_pert}) :
\beq
  \delta\Phi(x,y) = -\mu_5 F(x,y) + \delta\Phi_0(t,y) \label{dP_pert}
\eeq
Lastly the $G_{i5}$ equation, after substituting eqs.\ (\ref{G_pert}) and (\ref{dP_pert}),
reduces to
\beq
  \del_j (F' + 3n'F) =0	\label{i5_derived}
\eeq
The general solution to (\ref{i5_derived}) is
\beq
  F(x,y) = \hat{F}(x)e^{-3n(y)} + F_0(t,y) \label{F_pert}
\eeq
We find that the  $\hat{F}(x)$ terms in the solutions to the metric and 
scalar field perturbations, eqs. 
(\ref{G_pert},\,\ref{dP_pert},\,\ref{F_pert}),
vanish at linear order from the off-diagonal constraint equations and so $\hat{F}(x)$
is undetermined.  Therefore it can be consistently set to zero.

Now let us specialize the general ansatz to a form that is adapted
for the cosmological solutions we seek:
\bea
  ds^{2}&=& e^{2n(y)}\left[e^{2F_1( \varphi(t),y) 
  + 2F_2(\varphi(t),y)}\left( dt^{2}-a^{2}(t)\, d\mathbf{x}^{2}\right)
  -b_{0}^{2}\,e^{2\varphi \left( t\right) -4F_2(\varphi(t),y) }dy^{2}\right]   \label{metric_pert}\\
   F_1(\varphi,y) &=& n(y)f(\varphi(t))\,,\quad  F_2(\varphi,y)=f_2(\varphi(t))e^{(3+\varepsilon)(n_1-n(y))} \label{f2metric_pert} \\
  \Phi(t,y) &=&  \mu_5\left( kb_0y - \varphi(t)\right) +\delta\Phi(\varphi(t),y) \label{scalar_pert}
\eea
where $n_1=n(1)=-kb_0$ and $\varepsilon$ is an adjustable
parameter.\footnote{We will see that the choice $\epsilon = -1$
leads to small back-reaction of $f$, $f_2$ and $\delta\Phi$ while
allowing for large $\varphi$} 
It reduces to the previous static solution when the radion fluctuation
$\varphi(t)$ vanishes, if $f$, $f_2$ and $\delta\Phi$ also vanish. 
One can interpret  $f,f_2$ and $\delta\Phi$ as small back-reactions
induced on the metric and scalar field, respectively, by the (possibly
large) radion
fluctuation $\varphi$.
This ansatz does not provide an exact solution
to the equations of motion, but it is a good approximation at leading order in 
$f$, $f_2$ and $\delta\Phi$ as long as these quantities are small, in a sense to be 
specified.  We will find situations where it is possible 
to reliably consider large excursions of the radion during inflation, even 
superPlanckian ones.  It will be shown that the functions $f$, $f_2$ and $\delta\Phi$
can be algebraically derived from the brane potentials $V_0$ and $V_1$.
In turn they determine the potential for the radion in the 4D effective theory.
We will find that the radion potential derived from this ansatz is 
dominated by contributions from $f$ and its bulk scalar
counterpart (the part that is flat in the extra dimension).
The other parts of the perturbations, those with 
exponential $y$-dependence, turn out to be relevant solely
for the stabilization of the static solution.

\subsection{The $G_{05}$ equation and junction conditions}

To arrive at the functional form of the scalar perturbations we examine the off-diagonal Einstein equation.
Linearizing the left hand side we find  
\beq
  -\dot{\varphi}\left( 1+f-\frac{df}{d\varphi } - 3F_2 + \frac{dF_2}{d\varphi}\right) =\mu_5^{-2}
  \,\dot{\Phi} \Phi' \label{05_pert}
\eeq
This motivates us to parametrize $\delta\Phi(\varphi,y) = \delta\Phi_1+\delta\Phi_2$  where
\beq
 \mu_5^{-1}\delta\Phi_1(\varphi,y) =  g(\varphi)\, , \qquad\qquad 
  \mu_5^{-1}\delta\Phi_2(\varphi,y) = g_2(\varphi)e^{(3+\varepsilon)(n_1-n(y))} \label{scalar1_time}  
\eeq
This separation is useful, because it 
will be necessary to require that $f,f_2,g,g_2\ll 1$ but not necessarily that 
$\varphi\ll 1$.  
With this parametrization, the dominant ``1'' terms which are zeroth order
in the perturbations cancel as in our previous analysis and 
the linearized $G_{05}$ equation becomes
\beq
  -f +\frac{df}{d\varphi }-\frac{dg}{d\varphi}=  
  e^{(3+\varepsilon)k(y-b_0)}\left((1+\varepsilon)\frac{df_2}{d\varphi} +\frac{dg_2}{d\varphi} -(3+\varepsilon)(f_2 + g_2)\right)
    \label{05_reduced_new}
\eeq
It is satisfied everywhere in the bulk if
\bea
  \frac{df}{d\varphi}- \frac{dg}{d\varphi} &=& f\label{05_reduced}\\
  (1+\varepsilon)\frac{df_2}{d\varphi} +\frac{dg_2}{d\varphi} &=& (3+\varepsilon)(f_2+ g_2) \label{05_f2g2}
\eea
and has negligible $y$-dependence when 
\beq
  f_2,g_2 \ll \varphi \qquad \text{and}\qquad   f_2,g_2 \ll 1 \label{small_f2g2} 
\eeq 
In our numerical analysis we find eqs. (\ref{05_reduced},\,\ref{05_f2g2}) hold to a high degree of accuracy
when the back-reaction is small. Moreover, for the cases we study, eq. (\ref{05_f2g2}) will reduce to eq. (\ref{05_reduced}).

Now let us reconsider the junction conditions in the presence of the perturbations:
\bea
    2\mu_5^2k\,\left.\left(1+f-(1+\varepsilon)f_2e^{-(3+ \varepsilon)(n_i-n_1)}\right) \,e^{-n_i-\varphi }\right|_{y=y_i}&=& 
    \pm \left.V_{i}(\Phi_i +\delta\Phi)\right|_{y=y_i} 
    \quad  \label{junction1_pert} \\
    2\mu_5 k\,\left.\left( 1 + \left((3+\varepsilon)g_2+2f_2\right)e^{-(3+ \varepsilon)(n_i-n_1)}\right) \,e^{-n_i-\varphi }\right|_{y=y_i} &=&
    \pm \frac{\partial V_{i}}{\partial \Phi }\left.(\Phi_i + \delta\Phi)\right|_{y=y_i} 
    \quad \label{junction2_pert}
\eea
where now $\Phi_0 = -\mu_5\varphi $ and $\Phi_1=
\mu_5(kb_0-\varphi)$ take the place of the unperturbed scalar field
at $y=0$ and $y=1$.
These are algebraic equations that determine the back-reaction functions $f,g$ and $f_2,g_2$.
At low energies the functions all have a linear dependence on $\varphi$ 
and are related as $g = f$ and $g_2 = (1+\varepsilon)f_2$.
However, when $\varepsilon=-1$, $g_2$ vanishes at linear order in $\varphi$ so 
that the bulk scalar perturbations are effectively independent of $y$.
Even when $\varphi$ is large, this choice simplifies the subsequent analysis without affecting the radion dynamics,  
so we will adopt $\varepsilon=-1$ from now on.

The linearized jump conditions can be solved to first order in $g$ by Taylor-expanding $V_0'$ 
about $\Phi = - \mu_5\varphi$
at $y=0$:
\beq
	g \cong - g_2e^{2n_1} + {2k e^{-\varphi}\left(1+ (2f_2 + 2g_2)e^{2n_1}\right) - \mu_5^{-1}\,V'_0\over
	V''_0}
\label{geq_new}
\eeq
where the prime on potentials denotes $d \over d\Phi$.
Expanding next for $V_0$ in eq. (\ref{junction1_pert}) we find
\beq	
	f = -1 +  {e^{\varphi}\over 2 \mu_5^2 k}\left(
	V_0 + \mu_5 (g+g_2e^{2n_1}) V'_0 \right)
\label{feq_new}
\eeq
Similarly, at $y=1$, we expand  $V'_1$ about $\Phi= \Phi_1 - \mu_5\varphi$ and find for arbitrary stabilizing potentials using
(\ref{geq_new}) and (\ref{feq_new})
\beq
    f_2 = -g_2\left[\frac{1}{2} +{V_0''V_1''e^{n_1}e^{\varphi}(1-e^{2n_1})\over 4k( V_0''+V_1''e^{3n_1})}\right] 
    -\left[{\frac{1}{2}(V_0''+V_1''e^{n_1}) +\frac{e^{\varphi}}{4\mu_5k}(V_0''V_1'e^{n_1} -V_0'V_1''e^{n_1})
    \over V_0''+V_1''e^{3n_1}} \right]		\label{f2eq} 
\eeq
Then expanding for $V_1$ in eq. (\ref{junction1_pert})
\bea
    g_2 = 
    G_2^{-1}&
    \left[ 2ke^{-\varphi}\mu_5(V_0'+V_1'e^{n_1})
     +  \left[e^{2n_1}\left((V_0 +V_1e^{n_1})V_1''e^{n_1} 
	-(V_0' +V_1'e^{n_1})V_1'e^{n_1}\right) \right.\right. 
	\nonumber\\
    & + \left.\left. (V_0 +V_1e^{n_1})V_0'' -(V_0' +V_1'e^{n_1})V_0' \right]
    \left(1-e^{2n_1}\right)^{-1}\right] 
    \label{g2eq}
\eea
where
\beq
    G_2 = V_0'V_1''e^{3n_1} -V_1'e^{n_1}V_0''
\eeq
A fine-tuned choice of brane potentials such that
$V_0 = -V_1e^{n1}$ is tempting for 
simplifying the solutions since then $g_2 =f_2 =0$. However,
our study of the KK excitations (section \ref{KK_sect}) shows that
this choice is precluded by the presence of a tachyon instability.  
The expressions for the back-reaction functions may nevertheless be
simplified by neglecting the subdominant contributions involving extra powers of the warp factor:
\bea
  g &\cong& {2k e^{-\varphi} - \mu_5^{-1}\,V'_0\over
  V''_0} \label{geq} \\
  f &\cong& -1 + {e^{\varphi}\over 2 \mu_5^2 k}\left(
  V_0 + \mu_5g V'_0 \right) \label{feq} \\
  g_2 &\cong& -g -\frac{2ke^{-\varphi}\mu_5\left(1+f \right) + \mu_5^{-1}V_1e^{n_1}}{V_1'e^{n_1}} \label{g2eq} \\
  f_2 &\cong& -g_2\left( 1 +\frac{e^{\varphi}}{4k}V_1''e^{n_1} \right)
      - \frac12 -\frac{e^{\varphi}}{4k}\left(\mu_5^{-1}V_1'e^{n_1} + V_1''e^{n_1}g \right) \label{f2eq} 
\eea

Our results for $f,g$ and $f_2,g_2$ are derived
from the jump conditions and not from the Einstein equations, so
it is not obvious that they satisfy the relation (\ref{05_reduced_new}) 
from the $G_{05}$ equation. However, it is simple to show that this indeed is the
case, at linear order.
Differentiating the first jump condition (at $y=0$ or $y=1$) in (\ref{junction1_pert})
with respect to $\varphi$ using (\ref{scalar1_time}) gives
\bea
  \left.\frac{dV_i}{d\Phi}\right|_{y=y_i} &=& \left.\frac{d\varphi}{d\Phi}\frac{dV_i}{d\varphi}\right|_{y=y_i} 
    = \mu_5\left.\left(-1 + \frac{dg}{d\varphi} 
+ \frac{dg_2}{d\varphi}e^{-2(n-n_1)}\right)^{-1}\frac{dV_i}{d\varphi}\right|_{y=y_i}
\eea
Eliminating ${dV_i\over d\Phi}$ using the second jump condition (at
$y=0$ or $y=1$) in (\ref{junction2_pert}) and linearizing,  the
dominant ``1'' terms again balance and we arrive directly at
(\ref{05_reduced_new}) evaluated on the branes (which also guarantees
that (\ref{05_reduced_new}) is satisfied in the bulk since both sides
of the equation vanish). This is similar to
what was found in ref.\ \cite{CGK} where the linearized off-diagonal
Einstein equation, $G_{\mu 5}$ evaluated on the branes was shown to be
equivalent to one of the junction conditions of the metric so that no
new constraints arise.

For small $\varphi \ll 1$ we can Taylor-expand about the background and arrive at simple expressions
for the back-reaction functions
\bea
  g &=& f , \quad 
  g = \frac{-\sigma_{0} \left(\varphi-g_2e^{-2kb_0}\right)}{ 1 -\sigma_{0}} \simeq -\frac{\sigma_{0}\varphi}{ 1 -\sigma_{0}} \\
  g_2 &=& 0, \quad  	
  f_2 = \frac12\frac{\varphi(\sigma_{1} - \sigma_{0})}{\left(1 -\sigma_{0} 
  - e^{-2kb_0}(1 -\sigma_{1})\right) } \simeq   \frac12\frac{\varphi(\sigma_{1} - \sigma_{0})}{(1 -\sigma_{0})} 
\eea
where the jump conditions 
\beq
  \pm \mu_5^{-1} e^{n_i} V_{i}'(\Phi)\vert _{\Phi_i} = \pm e^{n_i} \mu_5^{-2} V_{i}(\Phi)\vert _{\Phi_i} = 2k   
\eeq
have motivated us to parametrize the brane potentials such that
\beq
  \pm \, e^{n_i} V_{i}''(\Phi)\vert _{\Phi_i} = 2k(1-\sigma_{i}) \label{BPS_shift}
\eeq
This has a convenient physical interpretation. When $\sigma_i = 0$ the 
radion is massless and the background solution will be BPS in the sense of
the solution generating technique utilized in ref.\ \cite{dewolfe}.
In general, the $\sigma_i$ depend on parameters of the brane 
potentials, and they 
determine the zero-mode masses for the radion and the bulk scalar field in terms of
the curvature scale $k$. 
The effects of various choices of $\sigma_i$ will be explored in section \ref{sect4}.

\subsection{The remaining Einstein equations}

So far we have succeeded in finding the back-reaction as a function of the
radion fluctuation $\varphi$, but we have not yet determined the dynamics
of $\varphi$.  This comes from solving the remaining Einstein equations.
They are difficult to solve exactly due to their
explicit dependence on $y$ through the warp factor $n(y)$ which 
also appears in the metric and scalar perturbations.
However, when we supplement the restrictions (\ref{small_f2g2}) on $f_2$ and $g_2$ 
by analogous ones for $f,\, g$,
\beq
  f,g\ll \varphi \qquad \text{and}\qquad f,g\ll 1 \label{small_fg}
\eeq
then it is possible to linearize to obtain some equations that have negligible $y$-dependence.
The radion will induce only a small back-reaction on the 4D slices of the 
5D spacetime relative to the static solution,
\beq
  ds_{4}^{2}\simeq e^{2n\left( y\right) }\left( dt^{2}-a^{2}d\mathbf{x}%
  ^{2}\right) \left( 1+2nf\left( \varphi \right) + 2f_2(\varphi)e^{-2(n-n_1)} +\cdots \right) 
\eeq
Similar to the off-diagonal equation (\ref{05_reduced_new}),
zeroth order terms vanish 
since these involve factoring out $\varphi$ dependence with the
background solutions. The remaining equations can then be
averaged over the extra dimension to find the radion dynamics:  
\beq
  \int_0^1 dy \sqrt{g}\,G_N^M = \kappa_5^2 \int_0^1 dy \sqrt{g}\,T_N^M  \label{EE_averaging}
\eeq
It was shown in reference \cite{CGRT} that the cosmology of the radion in the RS model can be obtained equivalently 
from the effective action or from averaging the Einstein equations as in (\ref{EE_averaging}).
In Section 4.1 we will construct the 4D effective action that leads to the bulk-averaged equations of motion. 

The bulk equations to be averaged are given in the Appendix, eqs.\ (\ref{00_gen})-(\ref{55_gen}).
Collecting $t$ derivatives to the right hand side,
Using eq.\ (\ref{bulk_pot_2}) for $V$  and 
(\ref{scalar_pert}, \ref{scalar1_time}) for $\Phi$,
linearizing eq. (\ref{EE_averaging}) with $\kappa_5^{2}=3\mu_5^{-2}$  
and integrating over $y$ we find
to first order in the back-reaction functions
\bea
  \left(\frac{\dot{a}}{a}\right)^2 + \frac{\dot{a}}{a}\dot{\varphi}\left(1+\Upsilon_1\right)
  -\dot{\varphi}^2 \left(\frac{1}{2} + \Upsilon_2 \right)
  =  3k^2e^{-2\varphi}\left(f - g -\Upsilon_5 \right)  \label{00_linear_up} \\
  2\frac{\ddot{a}}{a} + \left(\frac{\dot{a}}{a}\right)^2
    + \left(2\frac{\dot{a}}{a}\dot{\varphi} + \ddot{\varphi} \right)\left(1 + \Upsilon_1 \right)
    + \dot{\varphi}^2\left(\frac{5}{2} + \Upsilon_3\right) 
  = 9k^2e^{-2\varphi}\left(f - g -\Upsilon_5 \right) \label{ii_linear_up}\\
  \frac{\ddot{a}}{a} + \left(\frac{\dot{a}}{a}\right)^2 
  + \ddot{\varphi}\left(\frac{\Upsilon_1}{2} + \Upsilon_4'\right)
  + 3\frac{\dot{a}}{a}\dot{\varphi} \left(\frac{\Upsilon_1}{2} + \Upsilon_4\right) 
  + \frac{\dot{\varphi}^2}{2}\left(1+ {\Upsilon_2 + \Upsilon_3 \over 2} +2\Upsilon_4\right) \nonumber \\
  = k^2e^{-2 \varphi}\left(4f - 3g - \Upsilon_6\right)	\label{55_linear_up}
\eea
where primes denote $\frac{d}{d\varphi}$ and 
\bea
  \Upsilon_1 \equiv -\frac{2}{3}f'\left(1 - 3kb_0e^{-3kb_0}\Omega \right),	\quad 
  \Upsilon_2 \equiv -\frac{1}{2}\Upsilon_1 - g'- 3e^{-3kb_0}\left(f_2'+ g_2'\right)(1-e^{kb_0})\Omega\quad \\
  \Upsilon_3 \equiv -\Upsilon_2 - 4g' - 12e^{-3kb_0}\left(f_2' + g_2' \right)(1-e^{kb_0})\Omega, \quad
  \Upsilon_4 \equiv 3e^{-3kb_0}f_2'(1-e^{kb_0})\Omega\quad\\
  \Upsilon_5 \equiv g_2e^{-3kb_0}(1-e^{kb_0})\Omega , \quad \Upsilon_6 \equiv 3e^{-3kb_0}(2f_2+5g_2)(1-e^{kb_0})\Omega 
\quad
\eea
with
\beq
    \Omega = (1 - e^{-3kb_0})^{-1} 	\label{omega_int}
\eeq
All of the $\Upsilon_i$ terms are subdominant to other similar terms appearing in the equations, so
it is consistent to drop them and keep only the explicit $f$ and $g$ source terms (notice that
$\Upsilon_{5,6}$ are exponentially suppressed by powers of the warp factor).  This demonstrates our earlier assertion 
that
the radion dynamics are dominated by the Planck
brane potential, which determines $f$ and $g$, while $f_2$ and $g_2$ are needed for stabilization only.
The remaining Einstein equations then take the 
simple FRW-like form
\bea
  \left( \frac{\dot{a}}{a}\right) ^{2}+\frac{\dot{a}}{a}\dot{\varphi}-\frac{%
  \dot{\varphi}^{2}}{2}&=&3k^{2}e^{-2\varphi }\left( f-g\right) 
  \label{00_linear} \\
  2\frac{\ddot{a}}{a}+\left( \frac{\dot{a}}{a}\right) ^{2}+\ddot{\varphi}+2%
  \frac{\dot{a}}{a}\dot{\varphi}+\frac{5}{2}\dot{\varphi}^{2}
  &=& 9k^{2}e^{-2\varphi }\left( f-g\right)
  \label{ii_linear} \\
  \frac{\ddot{a}}{a}+\left( \frac{\dot{a}}{a}\right) ^{2}+\frac{\dot{\varphi}%
  ^{2}}{2}& =&k^{2}e^{-2\varphi }\left( 4f-3g\right) 
  \label{55_linear}
\eea
These equations are augmented by the linearized Klein-Gordon equation
\beq
  \ddot{\varphi}+3\frac{\dot{a}}{a}\dot{\varphi}+\dot{\varphi}^{2}
  =4k^{2}e^{-2\varphi }\left( f-\frac{3}{2}g\right)  \label{scalar_eq_linear}
\eeq
found at this level of approximation by integrating (\ref{scalar_eq_general}) or equivalently
by starting from the 4D effective action (\ref{eff_act_Jordan}) in Jordan frame which we derive
in the next section. 

Of course, not all four of the equations 
(\ref{00_linear}-\ref{scalar_eq_linear}) are independent.  
As was observed in ref.\  \cite{JCHF1}, the linear combination
of equations
(\ref{scalar_eq_linear})$-$(\ref{ii_linear}) + 2$\times$(\ref{55_linear})
is equivalent to eq.\ (\ref{00_linear}).  Furthermore, 
$d\over dt$(\ref{00_linear})$+H(3\times(\ref{00_linear})-(\ref{ii_linear}))/
\dot\varphi + (\ref{scalar_eq_linear})+(\ref{00_linear})$ is equivalent to eq.\ (\ref{55_linear}).
The two constraints arise as a consequence of the gauge symmetries, namely
reparametrizations of the $t$ and $y$ coordinates \cite{JCHF1},\cite{CGRT}.

\section{Inflation in the effective 4D theory}
\label{sect4}

It is not difficult to
derive the form of the effective action that gives rise
to the 4D equations of motion (\ref{00_linear})-(\ref{scalar_eq_linear}).
Starting from the 5D action (\ref{action}) and the metric (\ref{metric_general}) one
arrives at
\bea
  \begin{split}
  S = 2 \int_0^{1}d^{\,5}x\, \sqrt{g}  \left[\mu_5^2\left(- e^{-2N}(\dot{A}^2 + \dot{A}\dot{B}) +2e^{-2B}N'^2 \right)
  + \frac{1}{2}\del^{\mu}\Phi\del_{\mu}\Phi -V(\Phi) \right]\\
  -\int  d^{\,4}x \sqrt{g_4}\left. V_0\right|_{y=0} -\int d^{\,4}x\sqrt{g_4} \left. V_1\right|_{y=1}
  \end{split}	\label{5Daction2}
\eea
The boundary contributions from the compact extra dimension exactly cancel the extrinsic curvature terms and so do not
appear in this expression.
Substituting for the metric functions $A,B$ and $N$ in (\ref{metric_pert}), for
$V$ and $\Phi$ with (\ref{bulk_pot_2}) and (\ref{scalar_pert}), and using the jump condition (\ref{junction1_pert}) to rewrite
$V_i$ in terms of the perturbations one can linearize about $\Phi = \mu_5(kb_0y - \varphi)$.
After integrating over $y$ we arrive at the effective action 
\bea
  S = \frac{2\mu_5^2}{3k\Omega} \int a^3e^{\varphi} dt\left[ 
  - \left(\frac{\dot{a}}{a}\right)^2  - \frac{\dot{a}}{a}\dot{\varphi}\left(1 + \Upsilon_1 \right)
    + \dot{\varphi}^2\left(\frac{1}{2} + \Upsilon_2\right) 
  -3k^2e^{-2\varphi}\left(f -g - \Upsilon_5 \right)\right]  \label{4Daction1}
\eea
Neglecting the small $\Upsilon_i$ terms as before, we can write the 
simplified effective action in Jordan frame as
\beq
  S = {1\over 2\kappa_4^2}\int dt\, a^3\, e^{\varphi}\left( -6\left({\dot a\over a}
  +\frac12\dot\varphi\right)^2 + \frac{9}{2}\dot\varphi^2 - 2\kappa_4^2 V_{r,J}(\varphi)\right)
  \label{eff_act_Jordan}
\eeq
where $V_{r,J }= 9k^2\kappa_4^{-2} e^{-2\varphi}(f-g)$ and the 4D Newton's constant $\kappa_4^2 = 8\pi G$ is found from  
\beq
  \kappa_4^{-2} = 2\kappa_5^{-2}\int^{1}_0 b_0e^{-3kb_0y}dy = \frac{2\mu_5^2}{9k\Omega}
  \label{4D_Newton}
\eeq
It is straightforward to show that this action
implies the equations of motion (\ref{00_linear})-(\ref{scalar_eq_linear}) to linear order in $f,g$
when we make use of eq.\ (\ref{05_reduced}).

One can go to the Einstein frame by performing the 
Weyl transformation
\beq
a\left( t\right)  =\gamma \bar{a}\left(\tau\right),
\qquad
dt =\gamma\, d\tau,\qquad \gamma =e^{-\frac{\varphi }{2}}
\label{conf_trans}
\eeq
after which the action takes the form
\beq
  S = {1\over 2\kappa_4^2}\int d\tau\, \bar a^3\, \left( -6\left({\dot{\bar a}\over
  \bar a}\right)^2 + \frac92\dot\varphi^2 - 2\kappa_4^2 V_r(\varphi)\right)
  \label{eff_act_Einstein}
\eeq
where dots now denote ${d\over d\tau}$, and the radion no longer mixes with 
the scale factor.  The Einstein frame potential is 
\beq
	V_r = {9k^2\over \kappa_4^2} e^{-3\varphi}(f-g)
\label{radpot}
\eeq
and $\varphi$ is related to the 
canonically normalized radion field $\phi$ by
\beq
	\phi = {3\over \sqrt{2}}\,\kappa_4^{-1}\,\varphi = {3\over \sqrt{2}}\,m_p\,
\varphi \equiv \mu_4\,\varphi
\label{cannorm}
\eeq
where $m_p$ is the reduced Planck mass.
The Friedmann equation then takes the usual form
\beq
	\bar H^2 = {\kappa_4^2\over 3}\left(\frac12\dot\phi^2 + V_r\right)
\eeq
as does the Klein-Gordon equation
\beq
  \ddot{\phi} + 3\bar{H}\dot{\phi} = -\frac{dV_r}{d\phi}
\eeq
We will continue to do some of the subsequent analysis using the
dimensionless field $\varphi$ however, to avoid having to repeatedly write $\mu_4$.

\subsection{Radion stability}

To summarize the results to this point, 
we have shown that in the class of 5D warped models we are considering, the
radion potential is given by
\beq
	V_r(\phi) = 9 k^2 m_p^2\, e^{-3\phi/\mu_4}(f-g)
\label{radpot}
\eeq
where
the functions $f$ and $g$ are determined by the Planck brane potential
$V_0$ to leading order for small $f,g$, by eqs.\ (\ref{geq}, \ref{feq}).
The TeV brane potential $V_1$ has virtually no effect 
because of the suppression of $f_2$ and $g_2$ by powers of the warp factor
in eqs.\ (\ref{geq_new}-\ref{feq_new}).   

We are assuming that $\phi=0$ is a stable equilibrium point.   
Since $f=g= f'-g' =0$ at $\phi=0$
(see eq.\ (\ref{05_reduced})),
it is clear that $\phi=0$ is a critical point of the potential, and $V_r$
also vanishes at this point, leading to a Minkowski solution.  To compute the
radion mass, notice that  $f''-g'' = f'$.  Therefore
\beq
  m_{r}^{2}=\left. \frac{d^{2}\bar{V}_{r}}{d\phi ^{2}}\right\vert _{\phi=0}
  ={9 k^{2}m_p^2\over \mu_4^2}f'(0) = 2 k^2 f'(0) = 2 k^2 g'(0)
  \label{radion_mass1}
\eeq
Using eq.\ (\ref{geq_new}), we find that $g'(0) = 1 - 2k/V''_0(0)$. (Recall that
$g' = {dg\over d\varphi}$ and $V_0'' = {d^2V_0\over d\Phi^2}$.)

Consider for example the exponential potentials (\ref{brane_pot1}-\ref{brane_pot2}).
The radion mass is determined by $V''_0(0) = 2k[\alpha_0^2(1-\beta_0) -
\beta_0^2(1-\alpha_0)]/(\alpha_0-\beta_0)$.  Recalling our discussion of the
hierarchy problem, where it was pointed out that values of 
$\alpha_1$ and $\beta_1$ close to unity alleviate the need for a strong explicit 
hierarchy between the scales of the potentials
on the two branes, we are motivated to define
\beq
	\alpha_1 = 1 + \hat\alpha,\qquad \beta_1 = 1 + \hat\beta  \label{a1b1defs}
\eeq
and for consistency when comparing parameters
\beq
  \alpha_0 = 1 + \alpha,\qquad \beta_0 = 1 + \beta   \label{abdefs}
\eeq
Then it is straightforward to show that 
$g'(0) = \alpha\beta(1+\alpha\beta)/(1-\alpha\beta)$.
The radion mass squared is thus given by
\beq
	m_r^2 = -2 k^2 \alpha\beta\, {1+\alpha\beta\over 1-\alpha\beta}
\label{radmass}
\eeq
and there is a second zero mode whose mass has a similar expression in terms of the
$\hat\alpha$ and $\hat\beta$ parameters.
Hence $m_r^2$ is generically nonzero, and it is positive as long as 
$-1 < \alpha \beta < 0$ or $\alpha\beta > 1$. 
As for the remaining parameters we will show in the next subsection that 
$0 < \hat\alpha\hat\beta \ll  1$ is required in order to avoid a tachyon instability
in the second light state, and to address the hierarchy problem. 

We emphasize an important difference between our model and the RS model
supplemented by Goldberger-Wise stabilization of the radion. In the latter,
the radion mass is suppressed by the warp factor and so is naturally at the TeV
scale, whereas in ours, whose background geometry is different, it is natually
at the scale $k$; we must do one tuning of parameters, $\alpha\ll 1$, to make it small enough for
inflation.  This will be advantageous when we consider the
robustness of the radion-as-inflaton potential against quantum
corrections from standard model physics later on.

\subsection{Kaluza-Klein excitations}\label{KK_sect}

One might wonder  if it is sufficient to stabilize the radion 
without also considering the zero mode of the bulk scalar field.
As ref.\ \cite{CGK} has shown, there is only a single KK
tower for the two fields.  However, in our model there is 
a second zero mode in the tower, the sign of whose mass squared depends 
on the choice of brane potentials.  We will show that this state
can be made much heavier than the radion so that it is consistent
to neglect it along with the higher KK modes.   Our
main point is to show that the nonzero KK modes are not excited
during inflation, which if it happened would invalidate our effective
4D description.  We will find that these modes have masses of order $k$,
while the second zero mode hass mass $\sqrt{\sigma_1} k$, with $\sigma_i$ defined
in (\ref{BPS_shift}). The details of the calculations are left to Appendix \ref{appB}.

Following Csaki, Graesser and Kribs (CGK) \cite{CGK} we consider the spectrum of
perturbations around the background solution. 
The metric ansatz used to describe the scalar fluctuations
differs slightly from theirs since we are working with a $y$-coordinate in which
the background is conformally flat%
\bea
  ds^{2}&=&e^{2n(y) }\left[ e^{2F(x,y)}\eta _{\mu \nu }dx^{\mu}dx^{\nu }-e^{2G(x,y)}dy^{2}\right]  \label{metricKK} \\
  \Phi (x,y)&=& \Phi_b(y) +\delta\Phi(x,y) =\mu_5 kby +\delta\Phi(x,y)
  \label{scalarKK}
\eea
A similar analysis was subsequently done by Kofman, Martin and Peloso (KMP) \cite{Kofman:2004tk}. 
They generalized the CGK results by constructing a gauge invariant 
combination of scalar perturbations which ensures the hermiticity and orthogonality
of modes in the mass spectrum, defined here as
\beq
  v \equiv z \left( F + \frac{\delta \Phi}{\mu_{5}}\right) 
  \label{defv}
\eeq
with  
\beq
  z \equiv \sqrt{2} \frac{\Phi_{b}'}{n'} e^{\frac{3n}{2}} =
	     -\sqrt{2} \mu_5 e^{-\frac{3ky}{2}}
  \label{defz}
\eeq
$v$ is 
analogous to the Mukhanov-Sasaki variable \cite{mukhanov,sasaki} in 4D
inflationary cosmology. 
Expanding the action (\ref{action}) to second order and diagonalizing
in terms of the different scalars, one finds separable solutions 
$v(x,y) = \sum_{j}Q_j(x)\tilde v_j(y)$
such that the  quadratic action of the KK modes can be 
reduced to \cite{Kofman:2004tk}
\beq
  S = \sum_{j} C_j \int d^{\,4} x \; Q_j \left[ -\Box - m_j^2 \,
  \right] Q_j  \label{KKactions}
\eeq
with the normalization coefficients
\bea
  C_j &\equiv& \frac{1}{2}\int_0^{1} dy\,b_0 \,\tilde{v}_j^2 +  
  \frac{\mu_5^2}{k} \left.e^{3n}\tilde F_j^2\right|_{y_0=0}^{y_1=1} \notag \\ 
  &=&
  \frac{1}{2}\int_0^{1}dy\, b_0e^{3n} \,
  \left(\frac{1}{kb_0}\tilde F'_j- 3\tilde F_j\right)^2 +  
  \frac{\mu_5^2}{k} \left.e^{3n}\tilde F_j^2\right|_{y_0=0}^{y_1=1}
  \label{KKnorm} 
\eea
where $\tilde F_j$ is related to $\tilde v_j$ as $F$ is to $v$ in
(\ref{defv}).
The solutions are determined by the jump conditions
 (\ref{junction1_KK},\,\ref{junction2_KK}) 
and the two nondynamical constraints (\ref{KKgauge},\,\ref{mu5_KK})
originating from the linearized off-diagonal Einstein equations to determine $G$ and $\delta\Phi$.

To calculate the mass spectrum in our model
we incorporate the results of KMP and CGK except in the application of the boundary conditions.
Unlike the present work, refs.\ \cite{CGK},\cite{Kofman:2004tk}
assumed a convenient stiff potential limit for the stabilizing potentials that 
forces $\delta \Phi$ to vanish on the branes with the result that only the radion zero mode
appears with the infinite tower of KK excitations. Relaxing this assumption gives rise to
the bulk scalar zero mode.
One can also understand the two light modes as being the moduli of the 
positions of the two branes in the bulk \cite{brax2003cosmological}.

General solutions obtained from the action expanded at second order for $v_j$ are equivalent to those 
obtained for $F_j$ from the linearized Einstein equations \cite{Kofman:2004tk}. 
We prefer to solve for $F_j$ similarly to CGK since the boundary conditions and normalization look simpler
in this language.
The heavier KK excitations and the zero modes take the form%
\bea
  \tilde F_j &=& e^{-\frac{3}{2}kb_0y}\left(A_j \sin (\sqrt{\lambda}_j b_0y) + B_j \cos (\sqrt{\lambda}_j b_0y)\right) 
  \label{KKsol} \\
  \tilde F_{z} &=&  e^{-\frac{3}{2}kb_0y}\left(A_{z} e^{\sqrt{\lambda}_{z}b_0y} + B_{z} e^{-\sqrt{\lambda}_{z} b_0y}\right)
  \label{radsol} 
\eea
where 
\beq
  \lambda_j = m_j^2 -\frac{9k^2}{4} \quad, \quad \lambda_{z} = \frac{9k^2}{4} - m_{z}^2 \label{KKradlambda}
\eeq
for $4m_j^2 \geq 9k^2$ and $m_{z}^2 \ll k^2$ respectively. The index $j\ge0$ is an integer which labels the heavier 
KK modes, while $z=r,s$ stands for the two light (radion or bulk scalar) zero-modes.
The complete solutions are obtained by applying the linearized boundary conditions to determine the
ratio of integration constants 
and the mass.

The lightest KK excited state has $\lambda_0 = 0$ with $m_0^2 = \frac{9}{4}k^2$. 
Heavier modes must be found numerically 
except for some special cases which we display in Table \ref{tab:table1}.
%
\begin{table}[h]
\begin{center}
\begin{tabular}{l|l|l} 
$\qquad\quad\sigma_i$ & \, $m_j^2$ (KK modes) & \, $m_{z}^2$ (zero modes) \\[5pt]\hline 
& &  \\[-12pt]
$\sigma_0 <0 \,,\sigma_1>0 \,,$ \,  & \, $m_0^2 = \frac{9}{4}k^2$ &
\, $m_{r,s}^2 \simeq -2\sigma_0 k^2 , \, \sigma_1 k^2 $ \\
 $|\sigma_0|\ll|\sigma_1| \ll 1$ &\, $m_{j+1}^2 \simeq \frac{9}{4}k^2+ \left[\left(j + \frac{1}{2}\right)\frac{\pi}{b_0}\right]^2$ \,  & 
\\[5pt] \hline  
& &  \\[-12pt]
$\sigma_i =0$  & \, $\frac{9}{4}k^2 + \left(\frac{j\pi}{b_0}\right)^2 $ &\, 0 \\[5pt]
$|\sigma_i| \gg 1$ & \, $\frac{9}{4}k^2+ \left(\frac{j \pi}{b_0}\right)^2$ & \, $2k^2$ \\[5pt]
$\sigma_0 = \sigma_1 ,\, |\sigma_i| \ll 1$ \,  & \, $\frac{9}{4}k^2+ \left(\frac{j\pi}{b_0}\right)^2$& 
\, $m_{r,s}^2 \simeq -2\sigma_0 k^2 , \, \sigma_0 k^2 $ 
\end{tabular}
\caption{\label{tab:table1}
\textbf{A summary of the mass spectrum for various brane potentials
parametrized by $\sigma_i$}. The entries correspond to the general case where tachyons are absent,
followed by the massless radion, the stiff potential limit and fine-tuned brane potentials.}
\end{center} 
\end{table}
%
Recall from the discussion in the previous subsection that $\sigma_1=\hat\alpha\hat\beta \ll 1$ is needed to explain the hierarchy between the Planck
and TeV branes.
If $|\sigma_0| \ll |\sigma_1| \ll 1$ so as to achieve a large mass gap between the radion and the other modes,
then the KK masses are well-approximated by
\beq
  m_{j+1}^2 \cong \frac{9}{4}k^2+ k^2\left[\left(j + \frac{1}{2}\right)\frac{\pi}{kb_0}\right]^2 \label{KKmass_heavy}
\eeq
for $j\ge 0$.
Thus all KK mode masses are at least order $k$, independent of the precise details of the stabilizing potentials.

We previously showed that the radion mass depends on $V_0$, the potential on the Planck brane; see eq.\ 
(\ref{radion_mass1})). 
Similarly the bulk scalar zero mode mass depends on the TeV brane potential $V_1$.
In the case where the brane potentials are tuned such that  $|\sigma_i| \ll 1$ 
(eq.\ (\ref{BPS_shift})) the zero mode masses are approximately
\beq
  m_{r}^2 \cong -2\sigma_0k^2 \, , \, \sigma_1 k^2
  \label{radmass2}
\eeq
To avoid tachyonic instabilities we must impose $\sigma_0 < 0$ and $\sigma_1 > 0$.
This result agrees with the analysis of criteria regarding the presence of tachyons
for brane world models utilizing the Goldberger-Wise mechanism carried out in reference \cite{lesgourgues2004}. 

Comparing (\ref{radmass2}) with (\ref{radmass}), for the exponential brane potential
$V_0$ (\ref{brane_pot1}) which implies $\sigma_0 = \alpha\beta$, we corroborate 
at first order in $\sigma_0$
the more exact expression for the radion mass given in the previous subsection.
We will henceforth assume that $\sigma_1=\hat\alpha\hat\beta$ 
from $V_1$ (\ref{brane_pot2}) is large enough so that the bulk scalar zero mode is
much heavier than the radion:
\beq
    |\sigma_0| \ll {\sigma_1} \ll 1  \label{radmass_restrict}
\eeq
We will verify that $m_s \gg H$ (the Hubble scale during inflation) so that it is consistent to
ignore the bulk scalar zero mode during inflation.  The KK modes are much heavier and thus can also
be considered as frozen.

\subsection{Radion inflation: an explicit model}
\label{two-exp}
Let us now consider the full radion potential (valid to leading order
in $f,g$) corresponding to the brane potentials 
(\ref{brane_pot1}-\ref{brane_pot2}), with the restrictions 
(\ref{0brane_pot_param0}-\ref{0brane_pot_param1}) 
and the definitions (\ref{abdefs}).  Ignoring the small exponential
terms involving $f_2$ and $g_2$
we find that
\beq
	g = {(\alpha-\beta)e^{\beta\varphi} 
	+ \beta(1+\alpha)e^{(\beta-\alpha)\varphi}
	-\alpha(1+\beta)\over 
	\alpha(1+\beta)^2 - \beta(1+\alpha)^2
	e^{(\beta-\alpha)\varphi}}
\label{geq_inf}
\eeq
and the combination $(f-g)$ that enters the radion potential
(\ref{radpot}) is
\bea
	f-g &=& {1+g\over\alpha-\beta}\left[\beta-\alpha
	+\alpha(1+\beta)e^{-\beta\varphi} - 
	\beta(1+\alpha)e^{-\alpha\varphi}\right]\nonumber\\
	&+& g {\alpha\beta\over \alpha-\beta} \left(
	e^{-\beta\varphi}-e^{-\alpha\varphi} \right)
\eea
In the case we are interested in where $|\hat\alpha|,|\hat\beta|\ll 1$ in (\ref{a1b1defs}),
the expressions for $f_2$ and $g_2$ (\ref{g2eq},\,\ref{f2eq})
take the simple form
then
\bea
	f_2 &\simeq& -\frac12 f +g + \frac12\hat\alpha\hat\beta(\varphi-\varphi^2)  \label{f2eq2} \\ 
	g_2 &\simeq& f -g + \frac12\hat\alpha\hat\beta\varphi^2  	\label{g2eq2}
\eea
Since $\varphi\sim  1$ during inflation while $\hat\alpha\hat\beta\ll 1$, it follows that
if $f$ and $g$ are small then so are $f_2$ and $g_2$ and the consistency conditions
(\ref{small_f2g2}) are satisfied.

To design a suitable potential for inflation, we have the two dimensionless
parameters $\alpha,\beta$ and one dimensionful scale $k$ at our disposal, 
subject to the restriction  (\ref{radmass_restrict}) on $\sigma_0=\alpha\beta$.
In addition $\sigma_1 = \hat\alpha\hat\beta$ must satisfy (\ref{radmass_restrict})
and we must ensure that $f,g\ll 1$ for consistency of
the effective theory.  Notice from (\ref{geq_inf}) that if $|\alpha|\sim|\beta|$, then
generically $g\sim 1$ even if $|\alpha|,|\beta|\ll 1$.  However if 
\beq
	|\alpha| \ll |\beta|
\label{abineq}
\eeq
(or $\beta\ll\alpha$, giving qualitatively similar results) then $f$ and $g$ can
be made parametrically small: both vanish in the limit $\alpha\to 0$.  
This motivates our choice of parameter values.  We have also remarked that
the radion potential has a second degenerate (Minkowski) minimum because of the 
decompactification limit as $\varphi\to\infty$.  Thus the generic form of the
potential is that of a hilltop model with a barrier separating the minima
at $\varphi=0$ and $\varphi=\infty$.  
 We find that this barrier can be
made very flat when $\beta$ is close to $-3$, for $0< \alpha\ll 1$, making it possible
to get inflation from near the top of the barrier.  The shape of the potential is illustrated
for $\alpha=10^{-7}$ and a few values of $\beta$ near $-3$ in figure
\ref{potfig}.   Figure \ref{corrfig} shows that the condition
$f,g\ll 1$ is satisfied in the region $0 < \varphi < 3$ where
inflation starting from the top of the barrier might take place.  
Choosing $\hat\alpha = 10^{-2}$ and $\hat\beta = 6 \times 10^{-3}$ for the fiducial
values of $\alpha=10^{-7}$ and $\beta=-3$
is sufficient to ensure eqs. (\ref{f2eq2},\,\ref{g2eq2}) for $f_2$ and $g_2$ are
good approximations and that the mass squared of the second light zero mode is
much larger than that of the radion.  We will presently show that it also exceeds the Hubble
scale during inflation.
 
\DOUBLEFIGURE[t]{pot.eps, width=\hsize}{fg.eps, width=\hsize}{Radion
potential for $\alpha=10^{-7}$ and $\beta=-2.9,\, -3\,
-3.1$. \label{potfig}}{Logarithm of the functions $g$ and $f-g$
 that must be $\ll 1$
in the region of validity of the potential,
for the case $\alpha=10^{-7}$ and $\beta=-3$. \label{corrfig}}
\DOUBLEFIGURE[h]{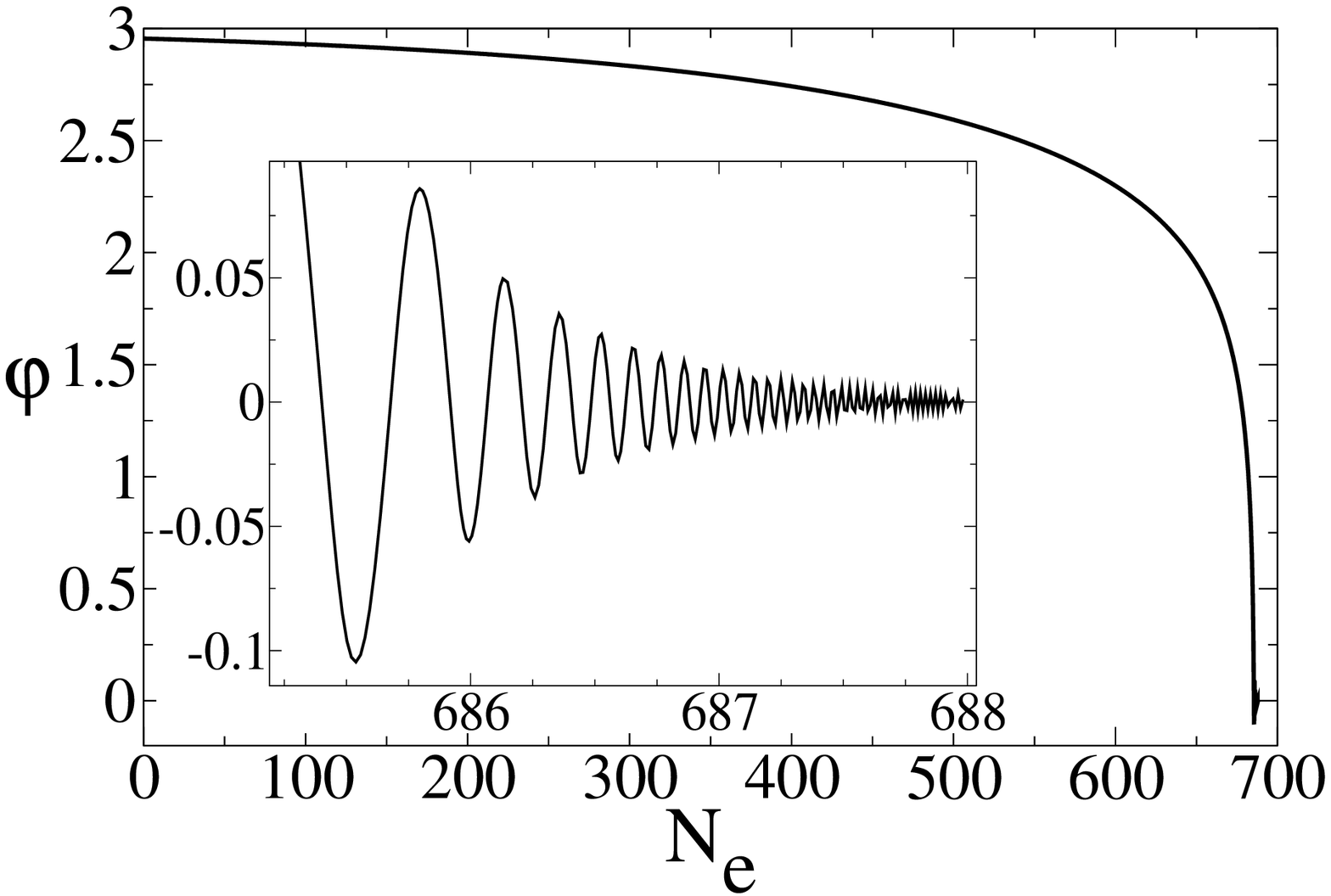, width=\hsize}{ns.eps, width=\hsize}
{$\varphi$ as a function of number of $e$-foldings $N$ for
$\alpha=10^{-7}$, $\beta=-3$ potential, starting $\Delta\varphi = 0.01$ from the
top of the barrier.  Inset shows end of inflation.\label{infsol}}{Spectral index as a function of
number of $e$-foldings until the end of inflation, corresponding
to fig.\ 3.\label{nsfig}}

We have numerically integrated the inflaton equations of motion
starting close to the top of the barrier.  Taking the number of 
$e$-foldings $N$ as the time variable, these can be written
in the first-order form,
\beq
	{d\varphi\over dN} = {\pi\over \mu_4 H};\qquad
	{d\pi\over dN} = -3\pi - {1\over \mu_4 H}{dV_r\over d\varphi}
\label{eoms}
\eeq
where $\pi = \dot\varphi$ is the canonical momentum and the Hubble parameter is
given by $H^2 = \frac16 \pi^2 + \frac13 V_r$ in Planck units.
(Recall that $\mu_4$ arises in the relation between the canonically normalized
radion $\phi$ and the dimensionless one $\varphi$.)
For $\alpha=10^{-7}$,
$\beta=-3$, the maximum of the potential is at $\varphi_m = 3.01$.
Starting at $\varphi=\varphi_m-\delta\varphi$ with $\delta\varphi=0.05$, for example, gives
$N_e= 685$ $e$-foldings of inflation.   $\varphi$ as a function of 
$N$ is shown in figure \ref{infsol}.  Although the total number of
$e$-foldings depends upon how small $\delta\varphi$ is taken to be, the
spectral index at the time of horizon crossing (nominally 60
$e$-foldings before the end of inflation) is insensitive to the
initial condition, as long as at least 60 $e$-foldings of inflation
occurred.  This behavior has been observed in a number of previous studies of
hilltop inflation models \cite{racetrack}-\cite{d3d7}.

To compute the power spectrum and spectral index 
we have used the prescription (see eq.\ (43) of ref.\ \cite{lyth})
\beq
	{\cal P} = {1\over 150\pi^2}\,{V_r\over m_p^4\epsilon},\qquad
	n_s = 1 + {d\ln{\cal P}\over dN}
\label{norm}
\eeq
in terms of the slow roll parameter $\epsilon = \frac12 m_p^2 (V_r'/V_r)^2 = 
\frac19 m_p^2 ({d V_r \over d\varphi} /V_r)^2$.  These are to be evaluated at horizon
crossing, approximately 60 $e$-foldings before the end of inflation.  
More precisely, we take the crossing of the scale relevant for the COBE
normalization of the power spectrum to be $N_{\rm\sss COBE} = 56.5$ \cite{better}, a number
that 
depends only logarithmically on the scale of inflation; since the spectrum
changes slowly with $N$, this is an adequate approximation for our purposes.

Fig.\ \ref{nsfig} shows the spectral index $n_s$ as a
function of $N_e-N$, the number of $e$-foldings until the end of 
inflation.  We see that $n_s=0.96$ at the scale of horizon crossing,
in agreement with the WMAP5 central value.  The normalization of the 
power spectrum at this point, $\sqrt{{\cal P}} = 2\times 10^{-5}$,
implies that the curvature scale is given by
$k = 0.015\, m_p$.   Eq.\ (\ref{radmass})
shows that the radion mass is 
\beq
m_r\cong \sqrt{6\alpha}k = 1.2\times 10^{-5}
\,m_p = 3\times 10^{13} {\rm\ GeV}.
\label{radmass13}
\eeq
{}From fig.\ \ref{potfig} the scale of inflation $m_I = V^{1/4}$ is given
by  $m^4_I \cong 3\times 10^{-8}\cdot 9 k^2 m_p^2$ in the flat
region of the potential, hence
\beq
	m_I =  2.7\times 10^{-3} m_p.
\eeq

The first slow roll parameter is 
$\epsilon = 1.1\times 10^{-4}$,  giving a small tensor-to-scalar ratio of
$r= 1.8\times 10^{-3}$.  This is related to the fact that the canonically normalized 
inflaton field does not change by an amount much greater than the Planck 
scale \cite{lyth-bound}; in our model $\Delta\phi = \frac{3}{\sqrt{2}} m_p \Delta\varphi
\cong 6 m_p$.  This is consistent with a refined version of the Lyth
bound \cite{mack}, $\Delta\phi \cong 6M_p\, r^{1/4}$ (where $M_p =
\sqrt{8\pi}\, m_p$).

We investigated the effect of varying the parameters $\alpha$ and
$\beta$ away from the fiducial values $\alpha = 10^{-7}$ and
$\beta=-3$.  Holding $\beta$ fixed and varying $\alpha$, we find that
for small $\alpha$ the spectral index asymptotes to its maximum value
of $n_s = 0.963$ (figure \ref{ns-alpha}).  Figure \ref{k-alpha} shows
that the  curvature scale $k$, determined by the normalization of
the power spectrum, follows a power-law relation $k\sim
\alpha^{-1/2}$ for very small values of  $\alpha$.  Recall that the
prefactor of the inflaton potential is proportional to $k^2 m_p^2$;
however the actual scale of inflation does not rise despite the
increase in $k$; the $\epsilon$ slow roll parameter asymptotes to 
$1.2\times 10^{-4}$ at the smallest values of $\alpha$,
which is nearly the same as its value at $\alpha=10^{-7}$.

\DOUBLEFIGURE[t]{ns-alpha-old.eps, width=\hsize}{k-alpha-old.eps, width=\hsize}
{Spectral index versus $\log_{10}\alpha$ for $\beta=-3$ model.
 \label{ns-alpha}}{Log of  curvature scale $k$ versus
$\log_{10}\alpha$ for $\beta=-3$ model.\label{k-alpha}}
\DOUBLEFIGURE[h]{ns-beta-new.eps, width=\hsize}{k-beta-new.eps, width=\hsize}
{Spectral index versus $\beta$ for $\alpha=10^{-7}$ model.
 \label{ns-beta}}{Log of curvature scale $k$ versus
$\beta$ for $\alpha=10^{-7}$ model.\label{k-beta}}

The corresponding results of varying $\beta$ at fixed $\alpha=10^{-7}$
are shown in figures
\ref{ns-beta} and \ref{k-beta}.  We see that it is possible to 
achieve a higher value of the spectral index than by
varying only $\alpha$; the maximum value is $n_s=0.99$ at
$\beta=-3.1$.

\subsection{Model with larger tensors}

In the previous model, constructed from brane potentials consisting of just
two exponentials, there was not enough freedom to find examples with a
tensor-to-scalar ratio $r$ exceeding $10^{-3}$.  However with the addition of
just one more exponential term, we can overcome this limitation. Consider the
Planck brane potential
\beq
V_{0} =\Lambda _{0}e^{\alpha _{0}\Phi /\mu _{5}}-\Delta _{0}e^{\beta
_{0}\Phi /\mu _{5}}-\Omega _{0}e^{\delta _{0}\Phi /\mu _{5}} 
\label{extended}
\eeq
The back-reaction functions $f$ and $g$ are given by
\bea
f &=&-1+{e^{\varphi}} \left( { \Lambda_0}\,{e^{-{ \alpha_0}\,\varphi}}-{ \delta_0}\,{e^{-{
 \beta_0}\,\varphi}}+{ \Omega_0}\,{e^{-{ \delta_0}\,\varphi}}+{ g}\, \left( {
 \Lambda_0}\,{ \alpha_0}\,{e^{-{ \alpha_0}\,\varphi}}-{ \delta_0}\,{ \beta_0}\,{e^{-{
 \beta_0}\,\varphi}}+{ \Omega_0}\,{ \delta_0}\,{e^{-{ \delta_0}\,\varphi}} \right) 
 \right) \nonumber\\
	g &=& -{\frac {{e^{-\varphi}}-{ \Lambda_0}\,{ \alpha_0}\,{e^{-{ \alpha_0}\,\varphi}}+{ \Delta_0
}\,{ \beta_0}\,{e^{-{ \beta_0}\,\varphi}}-{ \Omega_0}\,{ \Delta_0}\,{e^{-{ \Delta_0}\,
\varphi}}}{-{ \Lambda_0}\,{{ \alpha_0}}^{2}{e^{-{ \alpha_0}\,\varphi}}+{ \Delta_0}\,{{
 \beta_0}}^{2}{e^{-{ \beta_0}\,\varphi}}-{ \Omega_0}\,{{ \Delta_0}}^{2}{e^{-{ \Delta_0
}\,\varphi}}}}
\eea
which lead to the radion potential (\ref{radpot}), proportional to
$e^{-3\varphi}(f-g)$. 

Solving the unperturbed junction conditions similarly to eq.\ (%
\ref{0brane_pot_param0}-\ref{0brane_pot_param1}), we obtain
\bea
{\Lambda_{0}\over 2\mu_5^2 k} &=&{1\over {\beta-\alpha }}
\left( {\beta +\frac{\Omega _{0}}{2\mu
_{5}^{2}k}\,\left(\beta-\delta\right) }\right)\nonumber\\
{\Delta_{0}
\over 2\mu_5^2 k}&=&{1\over {\beta-\alpha }} 
\left( {\alpha +\frac{\Omega _{0}}{2\mu _{5}^{2}k}%
}\left({\alpha -\delta }\right)\right)
\label{0brane_pot_param0_2} 
\eea
where we have reparametrized $\delta_0=\delta +1$ 
similarly
to (\ref{abdefs}).  This model therefore has two parameters in addition to the
previous one, $\delta$ and $\Omega_0$, the latter of which we find convenient
to exchange for ${\Omega _{0}}/{2\mu _{5}^{2}k}\equiv h\alpha$. 
It will turn out that interesting values of 
${\Omega _{0}}/{2\mu _{5}^{2}k}$ are of order $\alpha$, so that 
$h$ is of order unity. 

\EPSFIGURE[b]{hpot.eps, width=0.5\hsize}
{Potential of extended model for  $h=0,1,2$, and $\alpha=1.5
\times 10^{-9}$, $\beta=-3$, $\delta=-2.9$.
 \label{hpot}}

\subsubsection{Analytic treatment}

We have seen that taking $\alpha\ll 1$ is needed to get acceptable
values of the spectral index.  The potential can be approximated by a
simpler expression in this regime, that allows us to make analytic
predictions for the tensor-to-scalar ratio $r$.  By further tunings
of parameters, we can achieve a potential that is nearly linear 
during inflation,\footnote{See ref.\ \cite{eva} for another recent example leading to a
linear potential, or ref.\ \cite{cliff} for other string-derived  
potentials supposed to be valid at Planckian field values.} leading to a large $\epsilon$ parameter and
$\eta\cong 0$.  This is the best case for getting a large tensor
signal from the model.  
Linear behavior in $\varphi$ occurs if some of the exponents in
$V_r$ are quite small.  Taking  $\delta$ is close to $-3$ realizes
this possibility.

First consider the small-$\alpha$ approximation to the potential, 
which can be written as%
\beq
V_{r}\cong9k^{2}m_{P}^{2}\alpha \left[ \frac{1-\delta h-3he^{-( \delta
+3) \varphi }-e^{-3\varphi }( 1+3\varphi -h( \delta
+3)) }{3+\alpha e^{3\varphi }( 4(1-\delta h)-3h(
1+\delta) ^{2}e^{-( \delta +3) \varphi }) }\right] 
\label{extend_rad_pot}
\eeq
This is a good approximation in the inflationary region,
 between $\varphi =0$ and the maximum of the potential.  
To see the turnover of the potential at large $\varphi$,  
it is important to keep the term of $O(\alpha)$ in the denominator; 
neglecting it
leads to a potential that levels out at large $\varphi$ instead of
having a maximum.  Next consider the case where $0 < \delta+3\ll 1$
so that the terms of order $e^{-3\varphi}$ in the numerator are 
subdominant.  This leads to 
\beq
V_{r}\cong9k^{2}m_{P}^{2}\alpha \left[ \frac{1-\delta h-3he^{-( \delta
+3) \varphi }}{3+ \alpha V_{\delta h}(\varphi)}\right]   
\label{extend_rad_pot_approx}
\eeq
The term $\alpha V_{\delta h}(\varphi)$ is $O(1)$ near the maximum of the
potential, but it becomes subdominant away from the maximum, and the
$\varphi$-dependence is dominated by the numerator.  In this
region, since $|\delta+3|\ll 1$, the dependence is in fact linear.
We can write $V_r\sim (1 -\xi + 3\xi\varphi)$, where $\xi=
h(3+\delta)$.  This potential has slow-roll parameters $\eta\cong 0$
and 
\beq
	\epsilon = \frac12 \left({m_p^2\over\mu_4^2}\right)
	\left( V_r'\over V\right)^2 = {\xi^2\over
(1-\xi+3\xi\varphi)^2}
\label{epsilon}
\eeq
Inflation ends when $\epsilon\sim 1$, {\it i.e.,}
\beq
	\varphi_e = {2\xi-1\over 3\xi}
\label{phi_end}
\eeq

It is straightforward to solve the slow-roll equation of motion
for the linear potential.  Setting $d\pi/dN=0$ in eq.\ (\ref{eoms}),
we have $d\varphi/dN = -(m_p/\mu_4)^{2} V'_r/V_r$. This gives a
simple quadratic function for $N(\varphi)$.  Using
(\ref{phi_end}), one can find
 $\varphi$ as a function the number of
$e$-foldings before the end of inflation, 
\beq
	\varphi(N) = \frac13\left(\sqrt{1+4N}+1 - \xi^{-1}\right)
\label{phiN}
\eeq
Notice that with $N\cong 60$ and $\xi\cong 1$, $\varphi(60)\cong
5$, in agreement with the position of the maximum of the potential
in fig.\ \ref{hpot}.  Substituting this into (\ref{epsilon}) we get
a simple result in which the dependence on the slope parameter
$\xi$ drops out completely:
\beq
	\epsilon = {1\over 4N+1} \cong 0.0044
\label{epsN}
\eeq
where we have taken $N=56.5$ to correspond to the COBE scale.  This gives a
tensor-to-scalar ratio of
\beq
	r = 16\epsilon = 0.07
\label{rlin}
\eeq
which is somewhat higher than the projected sensitivity of the
Planck experiment, $r_{\rm min} \cong 0.05$ \cite{planck}.  Moreover the spectral
index is
\beq
	n_s = 1 - 6\epsilon = 0.974
\label{nslin}
\eeq

The prediction that $r$ is independent of the parameters of the
potential in this regime depends on the assumption that the linear
behavior in the numerator of (\ref{extend_rad_pot_approx}) dominates
over the $\varphi$-dependence from $V_{\delta h}(\varphi)$ in the 
denominator.  In the next subsection we will quantify this by solving the equations of motion
numerically using the full potential.

Using eqs.\ (\ref{phiN}-\ref{epsN}), the power spectrum (\ref{norm}) 
in this model is $P=3(150\pi^2)^{-1}\alpha \xi k^2 m_p^2$ $(4N+1)^{3/2}$,
allowing us to determine $\sqrt{\alpha} k$ through the normalization:
\beq
	\sqrt{\alpha}k =7.6\, \xi^{-1/2} \times 10^{-6}\, m_p
\label{ak}
\eeq
The radion mass differs from that in (\ref{radmass13}) only by an extra
factor of $\sqrt{1+\xi}\sim 1$. 

\subsubsection{Numerical analysis}

Numerical analysis shows that the linear behavior of the potential
is fully achieved for somewhat smaller values of $\alpha\lsim
10^{-11}$ than we considered previously.  For larger values of 
$\alpha$, the linear shape is not fully realized and the tensor
ratio does not reach its maximum value of $r=0.07$.  This is
illustrated in figures \ref{tensor-h} and \ref{tensor-d}, which show
the dependence of $r$ and $n_s$ on $h$ and $\delta$ respectively,
when $\alpha = 1.5\times 10^{-9}$.  On the other hand, by decreasing
$\alpha$ to $1.5\times 10^{-12}$ and tuning $\delta$ closer to $-3$,
the predictions (\ref{rlin}) and (\ref{nslin}) based on the 
linear potential are borne out, both for $r$ and for $n_s$.  In all
cases, values of $h(3+\delta)$ of order unity are needed to get the
maximum tensor signal.

\DOUBLEFIGURE[h]{tensor-h.eps, width=\hsize}{tensor-d-new.eps, width=\hsize}
{Tensor ratio and $1-n_s$ versus $h$ for $\alpha=1.5
\times 10^{-9}$, $\beta=-3$, $\delta=-2.9$.
 \label{tensor-h}}{Tensor ratio and $1-n_s$ versus $\delta$ for
$\alpha=1.5
\times 10^{-9}$, $\beta=-3$, $h=3$ \label{tensor-d}}

\DOUBLEFIGURE[h]{h-smalla.eps, width=\hsize}{d-smalla.eps, width=\hsize}
{Same as fig.\ 10, but for $\alpha = 1.5\times 10^{-12}$
and $\delta = -2.99$.
 \label{h-smalls}}{Same as fig.\ 11, but for 
$\alpha = 1.5\times 10^{-12}$
and $h = 100$. \label{d-smalla}}

It is encouraging that the model is able to produce values of $r$
that exceed the minimum value of $r\cong 0.05$, which is estimated to
be at the threshold for detection by the Planck satellite
\cite{planck}. It is novel that we find such a regime in a hilltop
potential \cite{hilltop}, for which the tensor ratio is typically
unobservably small, $r< 0.002$.  We have checked that for the
examples shown here, the maximum value of $f,g$ in the inflationary
region is $f=0.02$, so the condition $f,g\ll 1$ is still satisfied
and the $O(f^2,g^2)$ corrections to our approximation for the
potential are under control. 

Consistent with the analytic prediction (\ref{ak}),  the curvature
scale $k$ starts to exceed $m_p$ in the large-tensor models that have
the smallest values of $\alpha$. For example with $1.5\times
10^{-11}$, $\delta = -2.99$ and $h=200$, we obtain $r=0.066$ and
$k=1.4 m_p$.  Examples with nearly as large tensors exist with $k <
m_p$, such as $\alpha=5\times 10^{-11}$, $\delta = -2.985$, $h=100$,
yielding $r=0.06$ and $k=0.8\, m_p$.  However it is the combination 
$\sqrt{\alpha}k$ that appears in the potential, and this remains
subPlanckian, so the fact that $k$ itself may exceed the Planck scale can
be regarded as merely an artifact of our parametrization.

\section{Coupling of radion to standard model}
\label{sect5}
An interesting feature of this model that distinguishes it from most other
models of inflation is that the couplings of the inflaton to the standard
model are exactly specified, since the radion is a component of the
higher-dimensional metric.  This enables us to address the details of reheating
in a way that is usually not possible without making additional assumptions.
Moreover, we can compute the quantum corrections to the radion potential to
check whether its exotic form is radiatively stable.

The most natural situation is that the SM fields are localized on the TeV
brane.  As an example, consider a Higgs field $H$.  Using the metric (\ref{metric_pert}) and 
the conformal transformation (\ref{conf_trans}), 
we find that in the Einstein frame,  its
coupling to $\varphi$ is 
\beq
{\cal L}_{\varphi,H} =  e^{-\varphi-2kb_0f + 2f_2}(\partial H)^2 
- e^{-2\varphi-4kb_0f + 4f_2} m_H^2 H^2 
\label{smcoupling}
\eeq
where we have renormalized the field and the mass to absorb warp factors
so that $m_H$ is the already-warped mass, $m_H\lsim$ TeV.
For small $\varphi$, $f(\varphi)\sim \alpha \varphi$ and $f_2(\varphi)\sim \hat\alpha\hat\beta \varphi$
which in the inflationary models we have considered is a negligible correction
to the leading $\varphi$ dependence  
in the exponents,  giving the usual result that the radion
couples to the trace of the stress-energy tensor.

\subsection{Reheating}
Using the conventional theory of perturbative reheating, one estimates the
reheat temperature as $T_r \sim g_*^{-1/4}(\Gamma M_p)^{1/2}$
where $\Gamma$ is the decay rate of the inflaton and $g_*$ the number of
relativistic degrees of freedom.  Using the interaction 
(\ref{smcoupling}) and recalling the relation (\ref{cannorm}) between $\varphi$
and the canonically normalized radion $\phi$ gives a rate of order $\Gamma
\sim g_*m_H^4/(16\pi m_p^2 m_r)$, since there are $\sim g_*$ additional degrees
of freedom besides the Higgs into which the radion can decay.  For $m_H\sim
$ 100 GeV, this leads to a reheat temperature that is too low even for
nucleosynthesis, $T_r\sim 10^{-3}$ eV.

However, there exists  a much more efficient channel for reheating,
by decay of the radion into gauge bosons.
Although the
radion does not couple to gauge bosons at tree level due to the tracelessness
of their stress energy tensor, the conformal anomaly induces a coupling at
one loop, of the form \cite{CGK}
\beq
	{\alpha\over 8\pi}\varphi F_{\mu\nu} F^{\mu\nu}
\eeq
where $\alpha$ is the fine structure constant.  An analogous term is present
for gluons and the electroweak vector bosons.  The decay rate due to
this operator is much faster than that due to (\ref{smcoupling}) since
the derivatives are of order $m_r/2$ as opposed to $m_H$.\footnote{we thank Neil
Barnaby for pointing out the possibility of perturbative decay 
into gauge bosons}\ \  The decay rate is approximately
\beq
	\Gamma \sim {\alpha^2 m_r^3\over 3^2\, 2^{13}\,\pi^3\, m_p^2}
\eeq
Using the radion mass (\ref{radmass13}) and $\alpha=0.1$ for QCD,
this gives a reheat temperature of order
\beq
	T_r \sim 10^7 {\rm\ GeV}
\eeq
which is high enough for electroweak baryogenesis, and low enough
to avoid the gravitino problem.

It is possible that reheating could be more efficient than indicated
by this perturbative estimate, due to parametric resonance by the same
coupling.   Production of massless particles can be a particularly
efficient form of preheating (see for example ref.\ \cite{Neil}). 

\subsection{Radiative corrections to inflaton potential}

The radion potentials we have derived have exotic shapes compared to simple
renormalizable potentials, and this has enabled us to produce distinctive 
signatures, like a larger tensor ratio than would be expected.  One should
always be concerned whether such peculiar potential shapes are radiatively
stable.  A second advantage of knowing the radion's couplings to matter is that
we can address this question quantitatively.  Consider the coupling
(\ref{smcoupling}) to a Higgs field.  To compute the contribution to the effective
potential for $\varphi$, we can take $\varphi$ to be constant and renormalize
$H$ to have a canonical kinetic term.  Then the contribution to the Coleman-Weinberg potential is
\beq
	\Delta V \sim e^{-2\varphi} {m_H^4\over 64\pi^2}
\eeq
Since $m_H^4 \ll k^2 m_p^2$ (the prefactor of the radion potential), this is a negligible correction.  Of course the
more relevant contributions come from integrating out particles
at some 
intermediate scale $m_{\rm int}$, if there is new physics above the TeV scale.  
Even then however we are safe as long as 
$m_{\rm int}^4 \ll k^2 m_p^2$.

One correction that does not appear in the Coleman-Weinberg potential
is the cosmological constant on the branes, a constant shift to the 
brane potentials (\ref{brane_pot1}-\ref{brane_pot2}).  Such terms can
be studied in the extended model with brane potential (\ref{extended})
in the limit where $\delta_0=0$ and by adding a similar term to the TeV brane.  
We have checked that our conclusions
based of the potential with two exponentials in \ref{two-exp} are not
very sensitive to the addition of the constant term.
Fig.\ \ref{ccpot} shows the effect of the constant term on the 
potential: for $h>0$, the flat region of the hilltop is shortened,
leading to a shorter period of inflation, while for $h<-1$ it
is destabilized.  The spectral index as a function of $h$ is shown for
the same potentials in fig.\ \ref{ccns}; values up to $h\cong 20$
are compatible with the $1\sigma$ WMAP5 allowed region,
$n_s=0.963\pm 0.015$.

\DOUBLEFIGURE[h]{cc_effect.eps, width=\hsize}{ns-cc-effect.eps, 
width=\hsize}
{Effect of constant addition to brane potential on inflaton potential,
parametrized  as eq.\ (4.17) with 
${\Omega _{0}}/{2\mu _{5}^{2}k}\equiv h\alpha$, for $\alpha=1.5\times 
10^{-9}$, $\beta=-3$, and a range of
values of $h=-10,\, 10,\dots,90$.\label{ccpot}}{Spectral index as a function of
$h$ for the potentials of fig.\ 14. \label{ccns}}

\section{Discussion}
\label{sect6}

We have explored a new framework for inflation, where the radion in a
warped 5D compactification is the inflaton.  Our approach differs from
the usual Randall-Sundrum setup in that we work in the linear dilaton
background (a bulk scalar that varies linearly with the extra
dimension) rather than AdS$_5$.  The bulk scalar stabilizes the radion
by the Goldberger-Wise mechanism.  Knowing its unperturbed solution
exactly gives us greater computational control over the back-reaction
when we perturb the radion away from the minimum of its potential.
We derived an 
approximate analytic formula for the radion potential that depends
mainly on the form of bulk scalar potential term on the Planck brane.

Similar to the RS model, the hierarchy between the weak and Planck scales
arises in our model without having to build it into the 5D Lagrangian.
One can check that for the dimensionless parameter values we needed for inflation, the
dimensionful brane potential parameters (\ref{0brane_pot_param0},\, \ref{0brane_pot_param1})
are dominated by the terms $\Lambda_0$ and $\Delta_1$, both of order
$\mu_5^2 k$ for a warp factor consistent with $kb_0 \sim 37$ needed to 
solve the hierarchy problem.
On the other hand,
the mass of the radion does not get warped down to the weak scale;
$m_r$ is fixed to be around $10^{13}$ GeV by the
normalization of the CMB power spectrum.  Moreover the couplings of 
the radion are not enhanced by the inverse warp factor; they remain
Planck-suppressed.  This tends to lead to a low reheat temperature, 
but also protects the inflaton potential from large radiative
corrections.  

Choosing a sum of two exponentials as the brane potentials,
we found that the radion potential suitable for inflation
was generically of the hilltop form, but (with some tuning of 
dimensionless couplings) having a shape that could be
much flatter than the generic  negative  quadratic form.  Expanding
the ansatz on the Planck brane to a sum of three exponential enabled us to find potentials
where the descent from the hilltop is linear and consistent with a 
large tensor contribution $r=0.07$, detectable by Planck.

An advantage of this model is that the couplings of the radion to
matter on the branes can be computed explicitly.  The radion couples
to the trace of the stress energy tensor of a given particle, and to
the one-loop conformal anomaly for states that are classically
conformally invariant.  We showed that this leads to reheating into
gauge bosons by perturbative decay, with a reheat temperature of order
$T_r\sim 10^7$ GeV. We did not explore the possibility of parametric
resonance, which might significantly increase $T_r$.  

A direction for further development is to see if our scenario can be
realized in a string theory compactification.  It is well-known that
warped-throat-like linear dilaton solutions arise in the near horizon
region of a stack of NS 5-branes \cite{CHS} (see also section 14.1 of
\cite{polchinski2}).  Moreover,  in the same way as the
Klebanov-Strassler throats \cite{KS} are smoothly capped versions of
anti-de Sitter throats near a stack of D3-branes, there exist known
smoothly capped throat-like linear dilaton solutions \cite{MN},
\cite{Chamseddine}. So far these throats and their compactifications
have not been as widely studied as the analogous Klebanov-Strassler
throat.

In ref.\ \cite{Greene} it was proposed that dynamically
weakening gravity at early times could provide the low-entropy initial
conditions that are needed for inflation to get started.  In the 
Jordan frame our effective 4D action (\ref{eff_act_Jordan}) has a 
time-dependent Newton's constant $\hat{G}(\phi) = Ge^{-\phi /\mu
_{4}}$ when the radion is displaced from its equilibrium value, which
could potentially realize this scenario.   In this frame, 4D gravity
is effectively weak when the radion is far from its stable minimum,
which is also the condition needed for inflation.   During inflation
$\hat{G}(\phi)$ grows to its normal strength and then oscillates about
this value as the radion decays.  It would be interesting to further
investigate the extent to which our model is compatible with this proposal.

Another feature of our model is that the mass spectrum of general
scalar perturbations, including the zero modes of the radion and bulk scalar field
and the infinite tower of KK excitations, is completely known in terms of $k$ 
and the parameters of the brane potentials.
All modes can be normalized and their exact 4D effective actions determined.
This presents an opportunity for constructing a model of assisted inflation
\cite{Liddle:1998jc} which we will explore elsewhere \cite{JCJT2}.

\paragraph*{\textbf{Acknowledgements:}}

We thank Andrew Frey, Alex Maloney and Johannes Walcher for
information about the linear dilaton background, Neil Barnaby for
insights about reheating and preheating, and Cliff Burgess
for discussions about radiative corrections to the brane potentials.

\appendix
\section{Bulk field equations}

\label{appA}
The ansatz of eq. (\ref{metric_general}) leads to the scalar field equation%
\beq
  \frac{d}{dt}\left( e^{-N+3A+B}\dot{\Phi}\right) -\frac{d}{dy}\left(
  e^{N+3A-B}\Phi ^{\prime }\right) +e^{N+3A+B}\left[ \frac{dV}{d\Phi }+\frac{
  dV_{0}}{d\Phi }\delta \left( y\right) +\frac{dV_{1}}{d\Phi }\delta \left(
  y-1\right) \right]  \label{scalar_eq_general}
\eeq
and after substituting into the Einstein equations, $G_{MN}=\kappa
^{2}T_{MN} $ we have
\begin{align}
  G_{0}^{0}& =3e^{-2N}\left( \dot{A}^{2}+\dot{A}\dot{B}\right) -3e^{-2B}\left(
  A^{\prime \prime }+2A^{\prime 2}-A^{\prime }B^{\prime }\right)
  \label{00_gen} \\
  G_{i}^{i}& =e^{-2N}\left( 2\ddot{A}+3\dot{A}^{2}+2\dot{A}\dot{B}-2\dot{A}
  \dot{N}-\dot{N}\dot{B}+\ddot{B}+\dot{B}^{2}\right)  \label{ii_gen} \\
  & -e^{-2B}\left( 2A^{\prime \prime }+3A^{\prime 2}+N^{\prime \prime
  }+N^{\prime 2}+2A^{\prime }N^{\prime }-2A^{\prime }B^{\prime }-N^{\prime
  }B^{\prime }\right)  \notag \\
  G_{5}^{5}& =3e^{-2N}\left( \ddot{A}+2\dot{A}^{2}-\dot{A}\dot{N}\right)
  -3e^{-2B}\left( A^{\prime 2}+A^{\prime }N^{\prime }\right)  \label{55_gen} \\
  G_{05}& =3\left( N^{\prime }\dot{A}+A^{\prime }\dot{B}-A'\dot{A}-\dot{A}
  ^{\prime }\right)  \label{05_gen}
\end{align}
in the bulk where the indices are $M,N=0,1,2,3,5$. The bulk contribution to
the stress energy tensor comes from 
\beq
  T_{MN}=g_{MN}V+\partial _{M}\Phi \partial _{N}\Phi -\frac{1}{2}\left(
  \partial ^{l}\Phi \partial _{l}\Phi \right) g_{MN}  \label{stress_energy}
\eeq
and we have the sources on the Planck and TeV branes: 
\begin{align}
  T_{N}^{M}& =e^{-B\left( 0,t\right) }\delta \left( y\right) \text{diag}\left(
  V_{0},V_{0},V_{0},V_{0},0\right)  \label{sources} \\
  & \;\;+e^{-B\left( 1,t\right) }\delta \left( y-1\right) \text{diag}\left(
  V_{1},V_{1},V_{1},V_{1},0\right)  \notag
\end{align}
We do not consider the effects of adding matter to the branes, since we are
primarily interested in the cosmology generated by the dynamical evolution
of the radion coupled to the bulk.

\section{Kaluza-Klein excitations}
\label{appB}
In this Appendix we provide details necessary for the computation
of the KK mass spectrum summarized in Section \ref{KK_sect}.
We closely follow the methods of CGK \cite{CGK} and KMP \cite{Kofman:2004tk}
and highlight differences in the 
zero mode solutions resulting from our different choice of
stabilizing potentials on the branes.  In particular, we have 
a second light mode due to not taking the stiff potential
limit.

Starting from the action of eq.\ (\ref{action})
we expand to second order in the scalar fluctuations (\ref{metricKK})
to arrive at \cite{Kofman:2004tk}  
\bea
  S &=& \frac{1}{2} \int d^{\,5} x \, v \left[ -\Box + \frac{d^2}{d y^2} -
  \frac{z''}{z} \right] v   \nonumber \\
  &&+ \int d^{\,5} x \, \partial_y \left[ \frac{\mu_{5}^2 e^{3n}}{n'}
  \eta^{\mu \nu} \partial_\mu F \partial_\nu F   
  + \frac{e^{3n}}{n'}  F \left( 4 n' \Phi_{b}' - {e^{2n} V'} \pm
  \frac{e^{n} \, V_{i}'' \Phi_{b}'}{2} \right) \delta \Phi \right]
  \label{action2nd}
\eea
where we define $v$ and $z$ as in eqs. (\ref{defv},\ref{defz}).
The eigenvalue equations for separable solutions to $v = \sum_{j}Q_j(x)\tilde v_j(y)$ 
are in Schr\"odinger form
\beq
  \left[ -\Box + \frac{d^2}{d \, y^2} - \frac{z''}{z} \right] v_j = 0 
  \label{eqv}
\eeq
while the second term of (\ref{action2nd}) is not yet in a simplified form.
To reach the compact form given by (\ref{KKactions}) in \cite{Kofman:2004tk}
we use an equivalent set of eigenvalue equations
involving $F =\sum_{j}Q_j(x)\tilde F_j(y)$ which 
can be derived from the linearized Einstein equations.

There are two nondynamical equations that we use to fix $G$ and $\delta\Phi$ in terms
of $F$. 
The first comes from recognizing that $\partial_{\mu}\partial_{\nu}$ terms at linear order should vanish
\cite{CGK} because there are no off-diagonal contributions in the stress energy tensor to source
such terms.
$\delta G_{\mu\nu} =0$ for these 4D components gives
\beq
  G = -2F   \label{KKgauge}
\eeq
which has already been applied in (\ref{action2nd}).
The second constraint comes from the 5D linearized off-diagonal 
Einstein equation, $\delta G_{\mu 5}=\kappa_{5}^{2}\delta T_{\mu 5}$. 
Inserting (\ref{KKgauge}) it can 
be integrated to find
\bea
  \delta \Phi =-\frac{\mu _{5}^2}{\Phi_b'}(F^{\prime } + 2n'F) 
  \label{mu5_KK}
\eea
where a $y$-dependent ``constant'' of integration has been set to zero to ensure
localization of the perturbations in $x$ (the large dimensions). With these constraints, the system of linearized Einstein equations 
is then reduced to a single dynamical equation
in the bulk found from the combination $\delta G_{\mu}^{\mu} -2\delta G_{5}^{5}$,
\bea
  F'' + 3n'F' = \Box F \label{Feqn}
\eea
where one also eliminates $V$ (eq. (\ref{bulk_pot_2}))
using the background solutions and the property
 $dV/d\Phi = 2\mu_5^{-1}V$ of our bulk potential.
Applying the definition 
(\ref{defv}) and the constraint (\ref{mu5_KK}) 
for eigenmodes $\square F_{j}=-m_{j}^{2}F_{j}$, eq. (\ref{Feqn}) is seen to be equivalent to 
that derived from the action involving $v_j$ (eq.\ (\ref{eqv})). 

To determine the mass spectrum we must apply the linearized boundary conditions.
At the Planck ($+$) and TeV ($-$) branes, 
\bea
  b_0^{-1}e^{-n} F'\vert_{y_{i}-\epsilon }^{y_{i}+\epsilon}
  &=&\pm 2F\mu _{5}^{-2} V_{i}(\Phi_i) \vert _{y_{i}}
  \mp \delta \Phi\mu_{5}^{-2} V_{i}'(\Phi_i) \vert _{y_{i}} 
  \label{junction1_KK} \\
  b_0^{-1}e^{-n}\delta \Phi'\vert_{y_{i}-\epsilon}^{y_{i}+\epsilon }
  &=& \mp 2F V_{i}'(\Phi_i)\vert _{y_{i}}
  \pm \delta \Phi V_{i}''(\Phi_i) \vert _{y_{i}}
  \label{junction2_KK}
\eea
where primes on the potentials denote $\frac{d}{d\Phi}$.
The ansatz for small fluctuations (\ref{metricKK}) makes
(\ref{mu5_KK}) redundant with the first boundary condition.
The second can be simplified using eq.\ (\ref{BPS_shift})
where we recall that the two would-be zero-modes get squared masses
proportional to $\sigma_1$ and $\sigma_2$. 
If we also rewrite eq.\ (\ref{Feqn})
in Schr\"odinger-like form
\beq
   \mathcal{F}_j'' - \frac{9}{4}n'^2\mathcal{F}_j = -m_j^2\mathcal{F}_j \label{Fheqn_KK}
\eeq
by defining
\beq
   F \equiv e^{-3n/2} \mathcal{F} \label{F_Schrod}
\eeq
so that
\beq
  \left( \Box + m_j^2 \right) Q_j = 0 ,\quad\quad
  \left( \frac{d^2}{d y^2} - \frac{9}{4}n'^2 + m_j^2 \right) \mathcal{\tilde F}_j
  = 0 \label{F_eigeqn}
\eeq 
and if we moreover apply eqs.\ (\ref{mu5_KK},\,\ref{Feqn}) at the boundaries,
 then (\ref{junction2_KK}) reduces to
\beq
   \mathcal{\tilde F}_j'|_{y=y_i} = \kappa_{ij}\,k\,b_0\,\mathcal{\tilde F}_j|_{y=y_i} \label{KKbndy}
\eeq
We have defined
\beq
  \kappa_{ij} \equiv \frac{m_j^2}{\sigma_{i}k^2} + \frac{1}{2} \label{def_kappa_ij}
\eeq
such that the index $i$ identifies the brane position ($y = y_i$) and $j$ is an integer corresponding
to the KK mode which we set to $z=r,s$ when treating the light radion and bulk scalar modes.

Solutions to (\ref{Fheqn_KK}), which we provide below, and consequently also to (\ref{eqv}) when inserted
back into the second order action (\ref{action2nd}), reduce
to the effective actions of the KK modes given by eq. (\ref{KKactions}). We
reproduce them here with the normalization coefficients given in terms of $\mathcal{\tilde F}_j$:
\bea
  S &=& \sum_{j} C_j \int d^4 x \; Q_j \left[ -\Box - m_j^2 \,
  \right] Q_j  \label{KKactionsB} \\
  C_j &\equiv&  \mu_5^2\int_0^{1} dy\,b_0 \left(\frac{\mathcal{\tilde F}'_j}{kb_0}- 
  \frac{3\mathcal{\tilde F}_j}{2}\right)^2 +  
  \left.\frac{\mathcal{\tilde F}_j^2}{k}\right|_{y_0=0}^{y_1=1}
  \label{KKnormB}
\eea
We have also substituted for the bulk potential (\ref{bulk_pot_2}), 
eliminated the $V_i''$ terms  in favour of perturbations using the boundary conditions (\ref{junction2_KK}),
and applied the constraint equation (\ref{mu5_KK}) for $\delta\Phi$.

The general solutions to the eigenvalue equations (\ref{F_eigeqn}) can be written
similarly to eqs. (\ref{KKsol},\,\ref{radsol}) for $\tilde F_j$
\bea
  \mathcal{\tilde F}_j &=& A_j \sin (\sqrt{\lambda}_j y) + B_j \cos (\sqrt{\lambda}_j y) \label{KKsolB} \\
  \mathcal{\tilde F}_{z} &=&  A_{z} e^{\sqrt{\lambda}_z y} + B_{z} e^{-\sqrt{\lambda}_{z} y} \label{radsolB} 
\eea
where as before
\beq
  \lambda_j = m_j^2 -\frac{9k^2}{4} \quad, \quad \lambda_{z} = \frac{9k^2}{4} - m_{z}^2 \label{KKradlambda}
\eeq
for $4m_j^2 \geq 9k^2$ and $m_{z}^2 \ll k^2$ respectively. 

Solving first for the KK modes,
at $y=y_0=0$ the constants are related by
\beq
  A_j = -\frac{B_j\,k\,\kappa_{0j}}{\sqrt{\lambda_j}} \label{KK_constants}
\eeq 
At $y=y_1=1$ we get a transcendental equation which is solved to find the masses
\beq
  k\sqrt{\lambda_j}\,(\kappa_{1,j} -\kappa_{0j})\cos(\sqrt{\lambda_j}\,b) =\left(k^2\kappa_{1j} - \lambda_j\right)
  \sin(\sqrt{\lambda_j}\,b)  \label{KKmass_eqn}
\eeq
The lightest KK excitation solution always gives $\lambda_0 = 0$ with $m_0^2 = \frac{9}{4}k^2$. 
The heavier modes must be found numerically except for the special cases of the stiff potential limit ($|\sigma_i| \gg 1$), 
a massless radion ($\sigma_i=0$), fine-tuned brane potentials ($\sigma_0 = \sigma_1$) or that of our model where
a large mass gap exists between the radion zero mode and the rest of the tower. In the former cases
one solves $\sin(\sqrt{\lambda_j}\,b) = 0$ to determine the masses. In our model 
when $|\sigma_0| \ll |\sigma_1| \ll 1$  we find $\cos(\sqrt{\lambda_j}\,b) \simeq 0$ and
\beq
  m_{j+1}^2 \simeq \frac{9}{4}k^2+ k^2\left[\left(j + \frac{1}{2}\right)\frac{\pi}{kb_0}\right]^2 \label{KKmass_heavyB}
\eeq

We next turn our attention to the zero-mode solutions.
Following the same reasoning using eq.\ (\ref{radsolB})
and (\ref{KKbndy}) at  $y=y_0=0$, 
we find the relation between the constants
\beq
  A_{z} = B_{z}\left(\frac{\sqrt{\lambda_{z}} + k\kappa_{0,z}}{\sqrt{\lambda_{z}} - k\kappa_{0,z}} \right)
  \label{rad_constants}
\eeq
At $y=y_1=1$ we get the equation that determines $m_{z}^2$,
\beq
  \lambda_{z} + k\sqrt{\lambda_{z}}(\kappa_{0{z}}-\kappa_{1{z}})\epsilon_{\lambda} 
  - k^2\kappa_{0{z}}\kappa_{1{z}} =0
  \label{zeromode_eqn}
\eeq
where
\beq
  \epsilon_{\lambda} =  \frac{1+e^{-2\sqrt{\lambda_{z}}kb_0}}{1-e^{-2\sqrt{\lambda_{z}}kb_0} }
\eeq
While eq.\ (\ref{zeromode_eqn}) does not yield a closed-form solution for the masses
in general, one can find analytic results in some approximations. 
Expanding $k\sqrt{\lambda_{z}} \simeq \frac{3k^2}{2} - \frac{m_{z}^2}{3}$ for $m_{z}^2 \ll k^2$ and ignoring the
small
exponential terms we can solve for the masses of the two light modes as
\beq
  m_{z}^2 \simeq -\frac{2\sigma_0k^2}{1-\frac{\sigma_0}{3}} \, , \, \frac{\sigma_1 k^2}{1+\frac{\sigma_1}{3}}
  \label{radmass2B}
\eeq
in agreement with eq.\ (\ref{radmass}).


\begin{thebibliography}{10}

\bibitem{HW}
  P.~Horava and E.~Witten,
  ``Eleven-Dimensional Supergravity on a Manifold with Boundary,''
  Nucl.\ Phys.\  B {\bf 475}, 94 (1996)
  [arXiv:hep-th/9603142].
  ``Heterotic and type I string dynamics from eleven dimensions,''
  Nucl.\ Phys.\  B {\bf 460}, 506 (1996)
  [arXiv:hep-th/9510209].


\bibitem{RS1}
  L.~Randall and R.~Sundrum,
  ``A large mass hierarchy from a small extra dimension,''
  Phys.\ Rev.\ Lett.\  {\bf 83}, 3370 (1999)
  [arXiv:hep-ph/9905221].


\bibitem{GW2}
  W.~D.~Goldberger and M.~B.~Wise,
  ``Modulus stabilization with bulk fields,''
  Phys.\ Rev.\ Lett.\  {\bf 83}, 4922 (1999)
  [arXiv:hep-ph/9907447].


\bibitem{rapid}
  N.~Arkani-Hamed, S.~Dimopoulos, N.~Kaloper and J.~March-Russell,
  ``Rapid asymmetric inflation and early cosmology in theories with
  sub-millimeter dimensions,''
  Nucl.\ Phys.\  B {\bf 567}, 189 (2000)
  [arXiv:hep-ph/9903224].


\bibitem{ADD}
  N.~Arkani-Hamed, S.~Dimopoulos and G.~R.~Dvali,
  ``The hierarchy problem and new dimensions at a millimeter,''
  Phys.\ Lett.\  B {\bf 429}, 263 (1998)
  [arXiv:hep-ph/9803315];
  I.~Antoniadis, N.~Arkani-Hamed, S.~Dimopoulos and G.~R.~Dvali,
  ``New dimensions at a millimeter to a Fermi and superstrings at a TeV,''
  Phys.\ Lett.\  B {\bf 436}, 257 (1998)
  [arXiv:hep-ph/9804398].


\bibitem{jcextra}
  J.~M.~Cline,
  ``Inflation from extra dimensions,''
  Phys.\ Rev.\  D {\bf 61}, 023513 (2000)
  [arXiv:hep-ph/9904495].

\bibitem{sundrum}
  R.~Sundrum and C.~M.~Wells,
  ``Warped Hybrid Inflation,''
  [arXiv:0909.3254 [hep-ph]].

\bibitem{Vilenkin}
  A.~Vilenkin,
  ``Topological inflation,''
  Phys.\ Rev.\ Lett.\  {\bf 72}, 3137 (1994)
  [arXiv:hep-th/9402085].

\bibitem{Linde:1994wt}
  A.~D.~Linde and D.~A.~Linde,
  ``Topological defects as seeds for eternal inflation,''
  Phys.\ Rev.\  D {\bf 50}, 2456 (1994)
  [arXiv:hep-th/9402115].


\bibitem{myers}
  R.~C.~Myers,
  ``New Dimensions For Old Strings,''
  Phys.\ Lett.\  B {\bf 199}, 371 (1987).


\bibitem{linear-dilaton}
  E.~Kiritsis, C.~Kounnas and D.~Lust,
  ``A Large class of new gravitational and axionic backgrounds for
  four-dimensional superstrings,''
  Int.\ J.\ Mod.\ Phys.\  A {\bf 9}, 1361 (1994)
  [arXiv:hep-th/9308124].

\bibitem{polchinski1}

J.\ Polchinski, ``String Theory, Vol. 1, An Introduction to the
Bosonic String,'' Cambridge University Press (2000); see eqs.\
(3.7.20) and (3.7.25).


\bibitem{dewolfe}
  O.~DeWolfe, D.~Z.~Freedman, S.~S.~Gubser and A.~Karch,
  ``Modeling the fifth dimension with scalars and gravity,''
  Phys.\ Rev.\  D {\bf 62}, 046008 (2000)
  [arXiv:hep-th/9909134].

\bibitem{BF}
  P.~Breitenlohner and D.~Z.~Freedman,
  ``Stability In Gauged Extended Supergravity,''
  Annals Phys.\  {\bf 144}, 249 (1982);
  ``Positive Energy In Anti-De Sitter Backgrounds And Gauged Extended
  Supergravity,''
  Phys.\ Lett.\  B {\bf 115}, 197 (1982).

\bibitem{KO} P. ~Kanti, S. ~Lee and K. ~Olive,
  ``Stable, time dependent, exact solutions for brane models
                        with a bulk scalar field'',
  Phys. Rev. D \textbf{67} 024037 (2003)
  [arXiv:hep-ph/0209036].

\bibitem{Kaloper:1999sm}
  N.~Kaloper,
  ``Bent domain walls as brane worlds,''
  Phys.\ Rev.\  {\bf D60}, 123506 (1999).
  [arXiv:hep-th/9905210].

\bibitem{JCHF} J. M. Cline and H. Firouzjahi, 
  ``Brane world cosmology of modulus stabilization with a
                        bulk scalar field,``
  Phys.Rev.\textbf{D} 64 023505
  (2001), [arXiv:hep-ph/0005235].

\bibitem{Langlois} D. Langlois and M Rodr\'{\i}guez-Mart\'{\i}nez, 
  ''Brane cosmology with a bulk scalar field,``
  Phys.\ Rev.\ D \textbf{64} 123507 (2001),
  [arXiv:hep-th/0106245].

\bibitem{KT1} K. Koyama and K. Takahashi, 
  ''Primordial fluctuations in bulk inflaton model,``
  Phys.\ Rev.\ D \textbf{67} 103503 (2003).
  [arXiv:hep-th/0301165].

\bibitem{KT2} K. Koyama and K. Takahashi, 
  ''Exactly solvable model for cosmological perturbations in
                        dilatonic brane worlds,''
  Phys.\ Rev.\ D \textbf{68} 103512 (2003),
  [arXiv:hep-th/0307073].

\bibitem{CGK}
  C.~Csaki, M.~L.~Graesser and G.~D.~Kribs,
  ``Radion dynamics and electroweak physics,''
  Phys.\ Rev.\  D {\bf 63}, 065002 (2001)
  [arXiv:hep-th/0008151].

\bibitem{JCHF1} J. M. Cline and H. Firouzjahi, 
  ``Five-dimensional warped cosmological solutions with
                        radius stabilization by a bulk scalar,''
  Phys.\ Lett.\ B {\bf 495} ,271 (2000) [arXiv:hep-th/0008185].


\bibitem{CGRT} C. Csaki, M. Graesser, L. Randall and J. Terning, 
  ``Cosmology of brane models with radion stabilization''
  Phys.Rev. D {\bf 62} 045015 (2000).
  [arXiv:hep-ph/9911406].

\bibitem{Kofman:2004tk}
  L.~Kofman, J.~Martin and M.~Peloso,
  ``Exact identification of the radion and its coupling to the observable
  sector,''
  Phys.\ Rev.\  D {\bf 70}, 085015 (2004)
  [arXiv:hep-ph/0401189].

\bibitem{mukhanov}
V.~F.~Mukhanov,
``Gravitational instability of the universe filled with a scalar field'',
  JETP Lett.\  {\bf 41}, 493 (1985)
  [Pisma Zh.\ Eksp.\ Teor.\ Fiz.\  {\bf 41}, 402 (1985)].

\bibitem{sasaki}
M.~Sasaki,
``Large Scale Quantum Fluctuations In The Inflationary Universe,''
Prog.\ Theor.\ Phys.\  {\bf 76}, 1036 (1986).

\bibitem{brax2003cosmological}
P. Brax, C. van de Bruck, A.C. Davis, and C.S. Rhodes,
  ``Cosmological evolution of brane world moduli,''
  Phys. \ Rev.\ D {\bf 67}, 023512 (2003)
  [arXiv: hep-th/0209158].

\bibitem{lesgourgues2004}
  J. Lesgourgues and L. Sorbo,
``Goldberger-Wise variations: Stabilizing brane models with a bulk scalar,''
  Phys.\ Rev.\  D {\bf 69}, 84010 (2004)
  [arXiv: hep-th/0310007].


\bibitem{racetrack}
  J.~J.~Blanco-Pillado {\it et al.},
  ``Racetrack inflation,''
  JHEP {\bf 0411}, 063 (2004)
  [arXiv:hep-th/0406230].


\bibitem{hilltop}
  L.~Boubekeur and D.~H.~Lyth,
  ``Hilltop inflation,''
  JCAP {\bf 0507}, 010 (2005)
  [arXiv:hep-ph/0502047].

\bibitem{better}
  J.~J.~Blanco-Pillado {\it et al.},
  ``Inflating in a better racetrack,''
  JHEP {\bf 0609}, 002 (2006)
  [arXiv:hep-th/0603129].

\bibitem{postma}
  Ph.~Brax, S.~C.~Davis and M.~Postma,
  ``The Robustness of $n_s < 0.95$ in Racetrack Inflation,''
  JCAP {\bf 0802}, 020 (2008)
  [arXiv:0712.0535 [hep-th]].

\bibitem{more-hilltop}
  K.~Kohri, C.~M.~Lin and D.~H.~Lyth,
  ``More hilltop inflation models,''
  JCAP {\bf 0712}, 004 (2007)
  [arXiv:0707.3826 [hep-ph]].

\bibitem{d3d7}
  C.~P.~Burgess, J.~M.~Cline and M.~Postma,
  ``Axionic D3-D7 Inflation,''
  JHEP {\bf 0903}, 058 (2009)
  [arXiv:0811.1503 [hep-th]].


\bibitem{lyth}
  D.~H.~Lyth and A.~Riotto,
  ``Particle physics models of inflation and the cosmological density
  perturbation,''
  Phys.\ Rept.\  {\bf 314}, 1 (1999)
  [arXiv:hep-ph/9807278].

\bibitem{lyth-bound}
  D.~H.~Lyth,
  ``What would we learn by detecting a gravitational wave signal in the  cosmic
  microwave background anisotropy?,''
  Phys.\ Rev.\ Lett.\  {\bf 78}, 1861 (1997)
  [arXiv:hep-ph/9606387].

\bibitem{mack}
  G.~Efstathiou and K.~J.~Mack,
  ``The Lyth Bound Revisited,''
  JCAP {\bf 0505}, 008 (2005)
  [arXiv:astro-ph/0503360].

\bibitem{eva}
  L.~McAllister, E.~Silverstein and A.~Westphal,
  ``Gravity Waves and Linear Inflation from Axion Monodromy,''
  [arXiv:0808.0706 [hep-th]].


\bibitem{cliff}
  M.~Cicoli, C.~P.~Burgess and F.~Quevedo,
  ``Fibre Inflation: Observable Gravity Waves from IIB String
  Compactifications,''
  JCAP {\bf 0903}, 013 (2009)
  [arXiv:0808.0691 [hep-th]].


\bibitem{planck}
The Planck Blue Book, 
www.rssd.esa.int/SA/PLANCK/docs/Bluebook-ESA-SCI(2005)1\_V2.pdf



\bibitem{Neil}
  N.~Barnaby, J.~R.~Bond, Z.~Huang and L.~Kofman,
  ``Preheating After Modular Inflation,''
  [arXiv:0909.0503 [hep-th]].

\bibitem{CHS}
  C.~G.~.~Callan, J.~A.~Harvey and A.~Strominger,
  ``Supersymmetric string solitons,''
  [arXiv:hep-th/9112030].


\bibitem{polchinski2}

J.\ Polchinski, ``String Theory, Vol. 2, Superstring Theory and Beyond,'' Cambridge 
University Press (2000). 



\bibitem{KS}
  I.~R.~Klebanov and M.~J.~Strassler,
  ``Supergravity and a confining gauge theory: Duality cascades and
  $\chi$SB-resolution of naked singularities,''
  JHEP {\bf 0008}, 052 (2000)
  [arXiv:hep-th/0007191].



\bibitem{MN}
  J.~M.~Maldacena and C.~Nunez,
  ``Towards the large N limit of pure N = 1 super Yang Mills,''
  Phys.\ Rev.\ Lett.\  {\bf 86}, 588 (2001)
  [arXiv:hep-th/0008001].

\bibitem{Chamseddine}
  A.~H.~Chamseddine and M.~S.~Volkov,
  ``Non-Abelian solitons in N = 4 gauged supergravity and leading order  string
  theory,''
  Phys.\ Rev.\  D {\bf 57}, 6242 (1998)
  [arXiv:hep-th/9711181];

  ``Non-Abelian BPS monopoles in N = 4 gauged supergravity,''
  Phys.\ Rev.\ Lett.\  {\bf 79}, 3343 (1997)
  [arXiv:hep-th/9707176].


\bibitem{Greene}
  B.~Greene, K.~Hinterbichler, S.~Judes and M.~K.~Parikh,
  ``Smooth Initial Conditions from Weak Gravity,''
  arXiv:0911.0693 [hep-th].

\bibitem{Liddle:1998jc} A. R. ~Liddle, A. ~Mazumdar, and F. E. ~Schunck, 
    ``Assisted inflation,''
      Phys.\ Rev.\ D{\bf 58}, 061301 (1998), [arXiv:astro-ph/9804177].


\bibitem{JCJT2} J.~M. ~Cline and J. ~Trudeau, \ in preparation. 


\end{thebibliography}
\end{document}